\pdfoutput=1 

\documentclass[a4paper,11pt]{article}

\usepackage{jheppub} 
\usepackage{xcolor,epsfig,amsmath,amssymb,subfigure,placeins}
\usepackage{graphicx,amsmath,bbm,bm,array}

\def\({\left(}
\def\){\right)}

\def\be{\begin{eqnarray}}
\def\ee{\end{eqnarray}}

\newcommand{\bea}{\begin{eqnarray}}
\newcommand{\eea}{\end{eqnarray}}
\def\({\left(}
\def\){\right)}

\renewcommand\Re{\operatorname{Re}}
\renewcommand\Im{\operatorname{Im}}

\usepackage[normalem]{ulem}  

\renewcommand\sout{\bgroup \color{red} \ULdepth=-.5ex \ULset}

\title{
Heavy quarkonia in a bulk viscous medium 
 }
\author[a]{Lata Thakur,}
\author[b]{Najmul Haque,}
\author[a,c]{and Yuji Hirono}

\affiliation[a]{
Asia Pacific Center for Theoretical Physics, \\
Pohang, Gyeongbuk 37673, Republic of Korea
}
\affiliation[b]{
School of Physical Sciences, National Institute of Science Education and Research, \\
HBNI, Jatni 752050, India
}
\affiliation[c]{
Department of Physics, POSTECH, \\
Pohang, Gyeongbuk 37673, Republic of Korea
}

\emailAdd{lata.thakur@apctp.org}
\emailAdd{nhaque@niser.ac.in}
\emailAdd{yuji.hirono@gmail.com}

\abstract{
We study the properties of heavy quarkonia 
in a quark gluon plasma in the presence of bulk viscous effects. 
Within the hard thermal loop approximation at one-loop, 
the dielectric permittivity of a quark gluon plasma is computed, 
where the bulk viscous effect enters 
through the deformation of 
the distribution functions of thermal quarks and gluons. 
Based on the modified dielectric permittivity, 
we compute the in-medium heavy quark potential, 
that includes non-pertubative string-like terms 
as well as the perturbative Coulombic term. 
We discuss how the bulk viscous effect modify the 
real and imaginary parts of the in-medium potential. 
Several prescriptions are examined as to how to 
include the string-like non-perturbative potentials. 
Using the deformed potential, 
we compute the wave functions, binding energies, 
and decay widths of heavy quarkonia in a bulk viscous medium, 
and study their sensitivity to the strength of the bulk viscous effect. 
An estimate of the melting temperatures 
is given. 
}


\begin{document}
\maketitle

\section{Introduction}

The relativistic heavy-ion collisions provide us  with a 
unique opportunity to experimentally study the 
strongly interacting matter in extreme conditions. 
The currently ongoing experimental programs
at the Relativistic Heavy Ion Collider (RHIC) at BNL and the Large Hadron Collider (LHC) at CERN
aim at revealing the properties of the quark gluon plasma (QGP), 
which is expected to appear at high temperatures. 
At sufficiently high temperatures, 
a QGP behaves as a weakly interacting gas of quarks and gluons, which can be understood using hard thermal loop (HTL) resummation~\cite{Weldon:1982aq,Braaten:1989mz,Frenkel:1989br,Braaten:1991gm}. 
Such a description has been  successful in describing the  thermodynamics of the QGP even close to the crossover temperature \cite{Andersen:1999fw,Andersen:2011sf,Haque:2014rua}.

Heavy quarkonium states have been a useful probe of the surrounding thermal medium. 
In the vacuum, 
they are reliably described in terms of non-relativistic potential models~\cite{Lucha:1991vn,Brambilla:2004jw} 
using the Cornell potential~\cite{Eichten:1974af, Eichten:1979ms}. 
A QGP medium exhibits the screening of static color-electric fields and that would result in the melting of heavy quarkonia, 
which was one of the first proposed signals of 
the formation of a QGP~\cite{Matsui:1986dk}. 
The potential models have been applied to 
the study of quarkonia at finite temperatures, 
the first of such works is done 
by Karsch, Mehr, and Satz~\cite{Karsch:1987pv}. 
The meson current correlators and quarkonium spectral functions have been calculated from potential models~\cite{Mocsy:2004bv,Wong:2004zr, Mocsy:2005qw,Cabrera:2006wh,Mocsy:2007jz,Alberico:2007rg,Mocsy:2008eg,Karsch:2000gi}
and  are compared to the first-principle lattice QCD calculations~\cite{Umeda:2002vr,Asakawa:2003re,Datta:2003ww,Aarts:2007pk,Jakovac:2006sf,Ding:2012sp,Ohno:2011zc,Larsen:2019zqv, Larsen:2019bwy}.
The appearance of the imaginary part of the potential due do the Landau damping~\cite{Laine:2006ns,Laine:2007qy,Burnier:2007qm,Beraudo:2007ky} and the break up of a color singlet bound state 
into a color octet quark-antiquark state via absorption of a thermal 
gluon is discussed~\cite{Brambilla:2008cx, Brambilla:2011sg, Brambilla:2013dpa},
which has further stimulated the study of complex heavy quarkonia potential
from thermal field theories 
\cite{Dumitru:2009fy,Petreczky:2010tk,Margotta:2011ta,Thakur:2013nia}
as well as from the lattice QCD~\cite{Rothkopf:2011db,Burnier:2012az,Burnier:2015nsa}.
See Ref.~\cite{Rothkopf:2019ipj} for a recent review.

The motivation for the current work is 
to understand how the effect of non-equilibrium nature of the fluid 
is imprinted on the properties of heavy quarkonia. 
For example, 
in the early time of a relativistic heavy-ion collision,
the longitudinal expansion is stronger than the radial expansion,
which would result in an anisotropy 
of the distribution functions of medium particles in the momentum space. 
The effect of such momentum-space anisotropies on quarkonia
has  been discussed in Refs.~\cite{Dumitru:2009ni,Dumitru:2009fy,Philipsen:2009wg,Dumitru:2010id,Margotta:2011ta,Thakur:2013nia,Thakur:2012eb,Strickland:2011aa,Strickland:2013uga,Strickland:2014pga}. 
The  presence of magnetic fields~\cite{Bonati:2015dka,Bonati:2016kxj,Marasinghe:2011bt,Alford:2013jva,Singh:2017nfa,Hasan:2017fmf,CS:2018jql,Kurian:2018dbn,Kurian:2020kct}
or non-zero fluid velocity~\cite{Escobedo:2011ie,Escobedo:2013tca,Thakur:2016cki,Avramis:2006em,Liu:2012zw,Ali-Akbari:2014vpa,Patra:2015qoa,Lafferty:2019jpr} also works as a source of anisotropy. 
Among such non-equilibrium situations, 
the role the bulk viscosity is gaining an increasing attention
in relation to the beam energy scan program \cite{Bzdak:2019pkr}, 
since the bulk viscous effect is expected to be enhanced 
as the system approaches a critical point~\cite{Kharzeev:2007wb,Karsch:2007jc,Moore:2008ws}.

The goal of this study is to test the sensitivity of quarkonia 
to the non-equilibrium nature of the fluid, 
in particular, the bulk viscous corrections. 
In Ref.~\cite{Du:2016wdx}, 
the color dielectric permittivity is computed 
in the presence of bulk viscous corrections 
based on the HTL-resummed gluon propagators. 
The bulk viscous effect enters through the deformation of the 
distribution functions of thermal particles. 
The perturbative HTL gluon propagators only gives rise to the Coulombic potential, 
but non-perturbative string-like contributions
have been observed in lattice QCD studies~\cite{Cheng:2008bs,Maezawa:2007fc,Andreev:2006eh}.
There has been several proposed prescriptions 
as to how to incorporate non-perturbative contributions in the potential~\cite{
Thakur:2012eb,Agotiya:2008ie,Guo:2019bwa,Burnier:2015nsa,Lafferty:2019jpr}. 
Among those is an approach based on the linear response theory: 
the modified string-like potential is obtained 
by modifying the linear potential using the HTL permittivity 
that entails the medium effect\footnote{
This approach  is taken in Refs.~\cite{Thakur:2013nia,Agotiya:2008ie,Thakur:2012eb} to study heavy quarkonia in an anisotropic medium. }. 
In this work, based on the modified dielectric permittivity in the presence 
of bulk viscous effect, 
we derive a complex heavy quark potential
in such environments. 
We examine several prescriptions for the introduction of non-perturbative part. 
We use the modified potential to solve the  Schr\"odinger equation 
and compute the deformed wave functions, 
binding energies, and decay widths of heavy quarkonia. 
We discuss how those physical properties are affected by the bulk viscous effect.

The rest of the article is structured as follows. 
In section~\ref{sec:permittivity}, 
we derive the dielectric permittivity of a thermal medium 
in the presence of bulk viscous corrections. 
In section~\ref{sec:potbulk}, 
we calculate the complex heavy quark potential 
based on the modified dielectric permittivity and 
discuss its properties. 
In section~\ref{sec:hq-property}, 
we show the effect of the bulk viscous corrections 
on the binding energies and decay widths of quarkonium states,
from which melting temperatures are estimated. 
Section~\ref{summary} is devoted to summary and discussions.

\section{
Color dielectric permittivity of
a bulk viscous medium 
}\label{sec:permittivity}

In order to compute the in-medium potential, 
we rely on the linear response theory, 
in which the in-medium properties are encoded in the color 
dielectric permittivity. 
We here review the derivation of the dielectric permittivity in the HTL approximation in the presence of the bulk viscous correction. 
When the system is away from thermal equilibrium, 
the fluctuation-dissipation theorem is violated, 
which leads to the existence of two different Debye masses. 
In the current situation, a modified fluctuation-dissipation theorem
is found to hold.

In the computation below, the non-equilibrium 
effect enters through the modification 
of the distribution function of thermal quarks and gluons, 
\begin{equation}
f({\bf k})=f_{0}(k)+\delta_{\rm noneq} f(\bf k)  ,
\end{equation}
where $ f_{0}(k) $ is the equilibrium distribution\footnote{
As a reference point, $f_0$, one may take a distribution of a non-thermal fixed point. 
}, 
and the second term is the non-equilibrium correction. 
In general, non-equilibrium corrections can be anisotropic. 
Such an anisotropy may be present at the early stage of heavy-ion collisions, where the longitudinal expansion is substantially stronger than the radial expansion. 
Certain types of the corrections can be regarded 
as viscous corrections when the anisotropy is weak. 
In this study, we discuss the effect of the bulk viscosity, 
and as $f_0$ we take the thermal equilibrium one, 
\begin{equation}
f({\bf k})=f_{0}(k)+\delta_{\rm  bulk} f(k), 
\end{equation}
where the correction $ \delta_{\rm bulk} f(k) $ is isotropic. 
The specific form of the correction is given later in Eq.~(\ref{eq:f-bulk})
\\

\subsection{
Computation of the dielectric permittivity 
}

Let us here 
derive the dielectric permittivity 
in the presence of bulk viscous corrections. 
For this, we compute the gluon self-energies and propagators 
in the presence of bulk viscous corrections. 
In the following, we review how to obtain the 
modified propagators as done in Ref.~\cite{Du:2016wdx}. 
The medium quarks are taken to be massless.

\subsubsection{Retarded propagator}

First, let us look at the retarded self-energy of gluons. 
In the following computations, we employ the Coulomb gauge. 
To evaluate the potential in the Coulomb gauge, we need the temporal component of the self energy 
$\Pi_{R}(P) \equiv \Pi_R^{00}(P)$\footnote{We denote a four momentum by a capital letter, $P = ( p^0, \mathbf p)$,
and $p\equiv |\mathbf p|$. 
}. 
In the HTL approximation, 
the one-loop contribution from $N_f$ quarks to $\Pi_R(P)$ is given  by 
\cite{Mrowczynski:2000ed,Du:2016wdx}
\begin{equation}
\Pi^{(q)}_{R}(P)=\frac{ 4\pi N_{f}g^2}{(2\pi)^4}\int kdk  d\Omega \(\frac{f^{+} ({\bf k})+f^{-} ({\mathbf  k})}{2}\)\frac{1-({\bf \hat{k}\cdot \hat{p}})^{2}}{\left({\bf \hat{k}\cdot \hat{p}}+\frac{p^{0}+i\epsilon}{p}\right)^{2}},
\label{PiR}
\end{equation}
where 
$\hat {\mathbf k} \equiv \mathbf k / k$
and $f^{\pm} (\bf k)$ are distribution functions for quarks/antiquarks. 
As long as the HTL approximation is valid, 
this expression is true even in non-equilibrium situations. 
In the  thermal equilibrium, the distribution functions are given by 
\begin{equation}
f^{\pm}_{0}(k)=\frac{1}{e^{(k\mp\mu)/T}+1},
\end{equation}
where $ \mu $ is the quark chemical potential. 
The contribution from the gluon loop has the same structure 
with the Fermi distribution replaced with the Bose one. 
Including the contribution from quark and gluon loops, 
the retarded self-energy in the equilibrium is written as 
\begin{equation}
\Pi^{\rm eq}_{R}(P) = m^{2}_{D}\left(\frac{p^{0}}{2p}\ln\frac{p^{0}+p+i\epsilon}{p^{0}-p+i\epsilon}-1\right) , 
\label{eq:pir-eq}
\end{equation}
where  $m_{D}$ is the Debye mass, 
\begin{eqnarray}
m^{2}_{D}=\frac{g^{2}T^{2}}{6}\left[2N_{c}+N_{f}\(1+\frac{3\tilde \mu^{2}}{\pi^{2}}\)\right] , 
\label{eq:debye-eq}
\end{eqnarray}
with $\tilde \mu \equiv \mu / T$. 

Now let us introduce the bulk viscous correction. 
We shall model the correction with the following form~\cite{Du:2016wdx}, 
\begin{equation}
\delta_{\rm bulk} f (k)=\(\frac{k}{T}\)^a \Phi\  f_{0}(k) \(1\pm f_{0}(k)\),
\label{eq:f-bulk}
\end{equation}
where $ a$ and $\Phi$ are constants and the $ +(-) $ sign is for Bose (Fermi) distribution. 
$ \Phi $ is a parameter 
proportional to the bulk viscous pressure (divided by ideal pressure). 
There are constraints for the parameter $a$. 
We need a condition $a>0$ 
so that there will be no IR divergence 
coming from the correction in the retarded self-energy of gluons 
(see Eq.~(\ref{PiR}) with $f$ replaced by the Bose distribution). 
Otherwise the dominant contribution does not come from $k \sim T$ 
and the HTL approximation becomes invalid. 
The same condition of the absence of IR divergence 
for the symmetric self-energy 
(gluon-loop version of Eq.~(\ref{eq:pis-quark})) 
leads to a stronger bound, $ a > 1/2 $, 
from the $\mathcal{O}(\Phi^2)$ contributions. 
 %
In addition, in order for 
the bulk viscous contribution to be non-negligible 
compared to the NLO corrections,
we need the condition $|\Phi| \gg g^2$. 
The bulk pressure $\delta_{\rm bulk} p$ is 
usually negative, which corresponds to $\Phi<0$, 
but the sign can be reversed in the presence of shear-bulk coupling \cite{Denicol:2014mca}. 
In the present study, we regard $\Phi$ as a parameter of either sign. 

In the presence of bulk correction (\ref{eq:f-bulk}), 
the retarded self-energy is modified as 
$\Pi_R  = \Pi^{\rm eq}_R  + \delta_{\rm bulk} \Pi_R$. 
The contribution from the quark loop, 
$ \delta_{\rm bulk}\Pi^{(q)}_{R}$, is given by
\begin{eqnarray}
\delta_{\rm bulk}\Pi^{(q)}_{R}(P)
&=&
\frac{  N_{f}g^2}{(2\pi)^3}\int kdk \left[\delta_{\rm bulk} f^{+}(k) + \delta_{\rm bulk} f^{-}(k) \right]
\int d\Omega \frac{1-({\bf \hat{k}\cdot \hat{p}})^{2}}{({\bf \hat{k}\cdot \hat{p}}+\frac{p_{0}+i\epsilon}{p})^{2}},\nonumber\\
&=&
\frac{ N_{f}g^2}{(2\pi)^3}\int kdk \(\frac{k}{T}\)^a \Phi\ \left[ f^{+}_{0}(k) \(1 - f^{+}_{0}(k)\) +f^{-}_{0}(k) \(1 - f^{-}_{0}(k)\) \right]\nonumber\\
&
\times&\left(\frac{p^{0}}{2p}\ln\frac{p^{0}+p+i\epsilon}{p^{0}-p+i\epsilon}-1\right).
\end{eqnarray}
The correction
does not affect the momentum dependence 
and just modify the Debye mass. 
Similarly, the contribution from the gluon loop can be computed. 
The total retarded self-energy including bulk correction can be written as~\cite{Du:2016wdx}
\begin{equation}
\begin{split} 
\Pi_{R}(P)
&=\Pi^{\rm eq}_{R}  + \delta_{\rm bulk} \Pi^{(q)}_R 
+\delta_{\rm bulk}\Pi^{(g)}_{R}
 \\
&=\widetilde {m}_{D,R}^2
\left(\frac{p^{0}}{2p}\ln\frac{p^{0}+p+i\epsilon}{p^{0}-p+i\epsilon}-1\right)  , 
\label{eq:pir-bulk}
\end{split}
\end{equation}
where 
$\widetilde {m}_{D,R}^2 = m^{2}_{D}+\delta m^{2}_{D,R}$
is the modified Debye mass. 
The correction is written as 
\begin{equation}
\delta m^{2}_{D,R}
=
\frac{g^2T^2}{6}\left[2N_{c}
c^{g}_{R}(a)\Phi
+N_{f}\(1+\frac{3 \tilde \mu^{2}}{\pi^{2}}\)
c^{q}_{R}(a,\tilde \mu)\Phi
 \right] . 
\label{mDRtot}
\end{equation}
Here, the dimensionless quantities
$ c^{q}_{R}(a,\tilde \mu) $ and $ c^{g}_{R}(a )$ are defined by  
\begin{eqnarray}
c^{q,g}_{R} = \frac{1}{\Phi} \frac{\int k dk\ \delta_{\mathrm{bulk}} f(k)}{\int k dk\ f_{\mathrm{0}}(k)}, 
\end{eqnarray}
where 
we take the Bose and Fermi distributions as $f_0$ for $c^g_R(a)$ and $c^q_R(a, \tilde \mu)$  respectively, 
and their explicit forms are  
\begin{eqnarray}
c^{q}_{R}(a, \tilde \mu) &=&
-\frac{6}{\pi^2+3 \tilde \mu^2 }
\ \Gamma(a+2)
\left[ {\rm Li}_{a+1}(-e^{\tilde \mu})
+ {\rm Li}_{a+1}(-e^{-\tilde \mu})\right], 
\\
c^{g}_{R}(a) &=&
\frac{6}{\pi^2}\Gamma(a+2)\zeta(a+1) ,
\end{eqnarray}
where ${\rm Li}_{n} (z)$ denotes the polylogarithm function. 
At the vanishing quark chemical potential $\mu=0$, 
the quark part $c^q_R$ simplify to 
\begin{equation}
	c^{q}_{R}(a, \tilde \mu=0 ) = \frac{12}{\pi^2} (1-2^{-a})  \Gamma(a+2)\zeta(a+1) . 
\end{equation}

We can compute the retarded propagator from the self-energy. 
In the Coulomb gauge, 
if the distribution function is isotropic, 
the temporal component of the resummed propagator\footnote{
Resummed propagators are indicated by 
characters with bars. 
}
, 
$\bar D_{R} (P)\equiv \bar D_{R} ^{00}(P)$ , 
is independent of the spatial components of 
the self-energy and propagators\footnote{
In the Coulomb gauge, 
the bare and resummed propagators satisfy the condition, 
\begin{equation}
p_i D^{0i}_R=0. 
\end{equation}
When the system is isotropic, like in the current case, 
we have $D^{0i}_R=0$. 
}. 
The Dyson-Schwinger equation reads 
\begin{equation}
{\bar{D}}_R = D_R + D_R \Pi_R \bar D_R, 
\end{equation}
where $D_R = 1/(p^2 + i \, {\rm sgn}(p_0) \epsilon )$
is the bare propagator. 
Using the $\Pi_R(P)$ obtained above, 
the temporal component 
of the resummed retarded propagator is written as
\begin{equation}
\bar D_{R} (P)=
 \frac{1}{p^{2}- \Pi_{R}} . 
\end{equation}
For the  computation of the potential, we need the  static limit $p_0  \to 0$ of the propagator. 
To the first order in $p_0$, it can be written as 
\begin{equation}
\bar D_R (P)
=
 \frac{1}{p^2+\widetilde m^2_{D, R}
 } 
-i \pi \frac{p_0} {2p} \frac{\widetilde m^2_{D, R}}{(p^2+\widetilde m^2_{D, R})^2} 
\Theta (p^2 - p^2_0)
+ O \left( p_0^2 \right) ,
\label{eq:dr}
\end{equation}
where $\Theta$ is the step function. 
The advanced propagator is given by the complex conjugate of the retarded one.

\subsubsection{Symmetric propagator}

The symmetric propagator $D_{S}(P)$ and 
the self-energy $\Pi_S (P)$ can be computed in a similar manner. 
The quark loop contribution to $\Pi_S (P)$ is written as
\begin{equation}
\Pi^{(q)}_{S}(P)=4iN_{f}g^2\pi^2\int \frac{k^{2}dk}{(2\pi)^{3}}\sum_{i=\pm} f^{i} (k)(f^{i} (k)-1)\frac{2}{p}\Theta(p^{2}-p_0^{2}). 
\label{eq:pis-quark}
\end{equation}
Adding the quark-loop and gluon-loop contributions, 
the equilibrium part of $\Pi_S(P)$ is written 
using  the Debye mass (\ref{eq:debye-eq}) as  
\begin{equation}
\Pi^{\rm eq}_{S}(P)=-2\pi i\, m_D^2 \frac{T}{p}\Theta(p^{2}-p_0^{2}). 
\label{eq:pis-eq}
\end{equation}
The total symmetric self-energy 
with the bulk viscous correction is again represented 
with the modified  Debye mass, 
\begin{equation}
\begin{split} 
\Pi_{S}(P)
&= 
\Pi^{\rm eq}_{S}
+ \delta_{\rm bulk} \Pi_{S} \\
&= 
-2\pi i
\widetilde m^2_{D, S} 
\frac{T}{p}\Theta(p^{2}-p_0^{2}), 
\label{eq:pis-bulk}
\end{split}
\end{equation}
where $\widetilde m^2_{D, S} =  m^2_D + \delta m^2_{D, S}$,
and the bulk viscous correction $\delta m^{2}_{D,S} $ is given 
to the first order in $\Phi$ by 
\begin{eqnarray}
\delta m^2_{D,S}=\frac{g^2 T^2}{6}\left(2N_{c}\,c^{g}_{S}(a)\,\Phi+N_{f}\(1+\frac{3 \tilde \mu^{2}}{\pi^{2}}\)\,c^{q}_{S}(a,\tilde \mu)\,\Phi\right), 
\label{mDStot}
\end{eqnarray}
where the dimensionless functions 
$ c^{q}_{S}(a,\tilde \mu) $ and $ c^{g}_{S}(a )$ are defined as 
\begin{eqnarray}
c^{q,g}_{S}
= \frac{1}{\Phi} 
\frac{
\int  dk \ k^2  \delta_{\mathrm{bulk}} f(k) (1  \pm 2 f_0 (k) )
}
{\int  dk\ k^2 
f_0 (k) ( 1 \pm f_0 (k) )
}, 
\end{eqnarray}
whose explicit forms are 
\begin{eqnarray}
c^{q}_{S}(a,\tilde \mu)&=&
-\frac{3}{ \pi^{2}+3\tilde{\mu}^{2} }
\Gamma(a+3)\left[{\rm Li}_{a+1}(-e^{-\tilde{\mu}})+{\rm Li}_{a+1}(-e^{\tilde{\mu}})\right], 
  \\
c^{g}_{S}(a)&=&\frac{3}{\pi^{2}}\Gamma(a+3)\zeta(a+1). 
\end{eqnarray}

Given the symmetric self-energy, we can compute the symmetric propagator. 
The temporal component of 
the resummed symmetric propagator in the Coulomb gauge 
satisfies the following Dyson-Schwinger equation 
\begin{equation}
\bar D_S
=
D_S
+ D_R \Pi_R \bar D_S 
+ D_S \Pi_A \bar D_A 
+ D_R \Pi_S \bar D_A, 
\label{eq:ds-s}
\end{equation}
Using the Dyson-Schwinger equation for the retarded and advanced propagators, 
$
\bar D_R = D_R + D_R \Pi_R \bar D_R, 
\,\,
\bar D_A = D_A + D_A \Pi_A \bar D_A, 
$
we can write Eq.~(\ref{eq:ds-s}) in the following form, 
\begin{equation}
(\bar D_R)^{-1} 
\bar D_S
(\bar D_A)^{-1}
= 
(D_R)^{-1} 
D_S
(D_A)^{-1}
+ \Pi_S . 
\label{eq:ds-s-2}
\end{equation}
The first term on the right hand side is in fact zero
(note that it is proportional to $p^4 \delta (p^2)$). 
Thus, the resummed symmetric propagator can be written 
as $\bar D_S = \bar D_R  \Pi_S \bar D_A$ and 
in $p_0 \to 0$ limit it is given by 
\begin{equation}
\bar D_S
=-\frac{2\pi i T  
\widetilde{m}_{D,S}^2
}{p(p^{2}+
\widetilde{m}_{D,R}^2
)^{2}}  
, 
\label{DS}
\end{equation}
where we have used Eqs.~(\ref{eq:pir-bulk}) and (\ref{eq:dr}).

\subsubsection{Dielectric permittivity}

The dielectric permittivity, $ \varepsilon(p) $, 
 is computed as
\begin{equation}
\varepsilon^{-1}(p)=\lim_{p^0  \to 0} {p^2} \bar D_{11} (P)~, 
\label{ephs}
\end{equation}
where $\bar D_{11}(P)$ is the longitudinal component
of the $11$-part of the resummed gluon propagator. 
Noting that 
\begin{equation}
\bar D_{11} =\frac{1}{2}\left( \bar D_R  + \bar D_A +\bar D_S  \right) ,
\label{eq:d11}
\end{equation}
and using Eq.~(\ref{eq:dr}) and Eq.~(\ref{DS}), 
we obtain the expression for the dielectric permittivity 
in the presence of bulk viscous correction, 
\begin{equation}
\varepsilon^{-1}(p) = 
\frac{p^2}{p^2 + \widetilde{m}^2_{D,R}}
 - i  
 \frac{\pi T p  \, 
\widetilde{m}^2_{D,S}  
}{ (p^2 + \widetilde{m}^2_{D,R})^2} . 
\label{eq:epsilon}
\end{equation}
In the limit of vanishing bulk correction, 
both of 
$\widetilde{m}_{D,R}$ and $\widetilde{m}_{D,S}$ 
approaches $m_D$ and  the equilibrium expression is reproduced. 
The dielectric permittivity (\ref{eq:epsilon}) will be used 
in computing the in-medium heavy quarkonia potential.

\subsection{
Two Debye masses and a modified fluctuation-dissipation theorem
}

We have learned that the effects of the bulk viscous correction
are incorporated in two different Debye masses, 
$\widetilde m^2_{D, R}$ and $\widetilde m^2_{D, S}$, 
that characterize 
the retarded (advanced) propagators and the symmetric propagator. 
The two Debye masses 
are functions of the bulk viscous correction parameter $\Phi$. 
We have obtained the expression of the modified dielectric permittivity
(\ref{eq:epsilon}), which is the main result of this section. 

\begin{figure}[tb]
\centering
		\includegraphics[width=10cm]{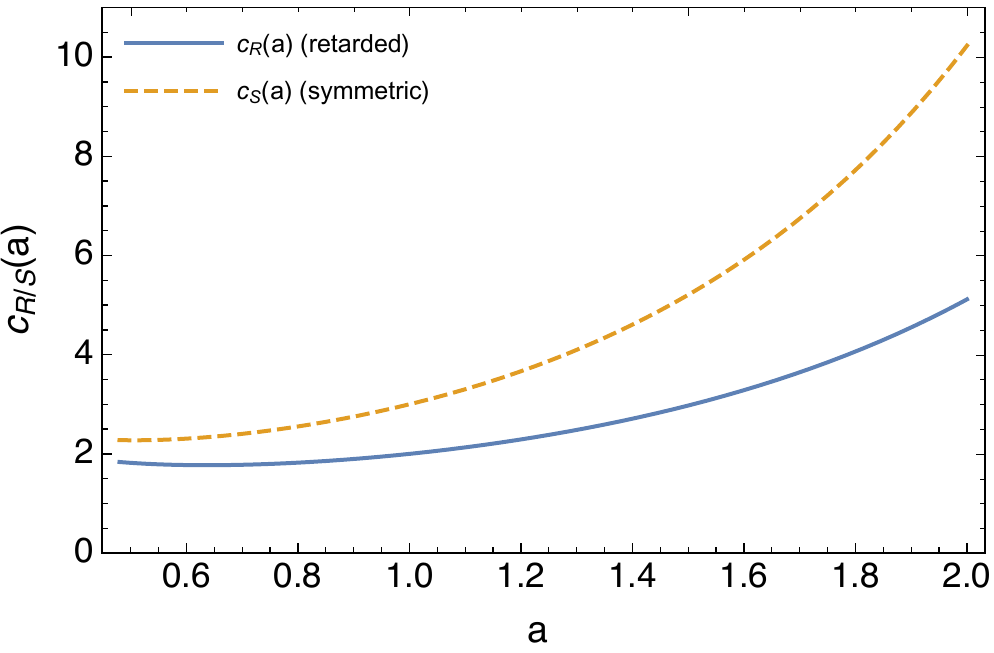}
\caption{
$a$-dependence of the functions $c_S(a)$ and $c_R(a)$, 
that are the slopes  of $\widetilde{m}^2_{R(S)} / m^2_D$ 
as a function of $\Phi$. 
The chemical potential $\mu$ is set to zero. 
}
\label{fig:dmsqdPhi}
\end{figure}

Let us show the behaviors of the modified Debye masses. 
The bulk viscous correction can be written as 
\begin{equation}
\frac{\widetilde m^2_{D, R(S)}}{m^2_D}
=
 1 + c_{R(S)}(a,\tilde \mu ) \Phi . 
\end{equation}
The expression of $c_R$ and $c_S$ follows from 
Eqs.~(\ref{mDRtot}) and (\ref{mDStot}). 
In Fig.~\ref{fig:dmsqdPhi}, we show the linear coefficients $c_R(a,\tilde \mu)$
and $c_S(a,\tilde \mu)$ as a function of $a$. 
Those coefficients are positive in the region of $a$ considered here. 
Therefore, the bulk viscous correction effectively
increases the Debye mass for $\Phi>0$.

In the absence of bulk viscous corrections, 
namely in the thermal equilibrium,
the fluctuation-dissipation theorem (FDT) holds, 
\begin{equation}
\bar D_S (P) = (1 + 2 f_0 (p_0)) {\rm sgn}(p_0)
\left(\bar D_R - \bar D_A \right) 
= 
\frac{2T}{p_0} {\rm sgn}(p_0) 
( \bar D_R - \bar D_A ) +  O( p_0 ) , 
\label{eq:fdt}
\end{equation}
and this ensures that 
the two masses are equal in the absence of bulk viscous corrections. 
The FDT is violated in non-equilibrium, 
and $\widetilde m_{D, R}$ and $\widetilde m_{D, S}$ can be different. 
Thus, the difference quantifies the extent of violation of the FDT. 
In the current situation, in fact, a modified version of FDT holds \cite{Carrington:1997sq}, 
\begin{equation}
\bar D_S (P) = \frac{2T \lambda }{p_0} {\rm sgn}(p_0) ( \bar D_R - \bar D_A ) +  O( p_0 ), 
\label{eq:mfdt}
\end{equation}
or equivalently 
\begin{equation}
\Pi_S (P)= \frac{2T \lambda }{p_0} {\rm sgn}(p_0) ( \Pi_R - \Pi_A ) +  O( p_0 ), 
\label{eq:mfdt-pi}
\end{equation}
where we have defined a parameter 
\begin{equation}
\lambda = 
\lambda(a, \Phi, \tilde \mu)
 \equiv 
\frac{ \widetilde m^2_{D, S} } { \widetilde m^2_{D, R} } 
= 
\frac{  1 + c_S \Phi }{ 1 + c_R \Phi }
\label{eq:def-lambda}
\end{equation}
The modified FDT (\ref{eq:mfdt}) holds
in the HTL approximation at one-loop 
and when the distribution function is spherically symmetric. 
As can be seen in Fig.~\ref{fig:dmsqdPhi}, the symmetric Debye mass is larger than the retarded one for $\Phi>0$, so $\lambda>1$ in this case.

\section{In-medium potentials in the presence of bulk viscous corrections}\label{sec:potbulk}

In this section, we study how 
the heavy quarkonia potential is modified 
in the presence of bulk viscous correction. 

A heavy-quark  potential can be obtained 
by the Fourier transform of the static gluon propagator. 
The propagators in the HTL perturbation theory 
results in the screened Coulombic potential. 
This potential will be dominant in the high temperature limit, 
but it does not account for the non-perturbative string-like part, which is responsible for the confinement. 
At zero or low temperature, 
many  studies has confirmed that the so-called Cornell potential
that consists of a Coulombic and string-like parts 
explain the properties of heavy  quarkonia very well. 
Modeling of non-perturbative effect 
is important in understanding the ``melting" of quarkonia 
near the crossover temperature $T_c$.

Given the medium property, how to incorporate it 
to modify the string-like contribution for both of the real and imaginary parts is not unique. 
There has been a number of proposals as to how to parametrize the 
in-medium potentials. 
Here, we shall discuss 
the prescriptions discussed in  Refs.~\cite{Guo:2019bwa, Lafferty:2019jpr} 
and 
we extend those formulations to introduce 
the non-equilibrium bulk viscous corrections. 
We examine how the real and imaginary part of the in-medium potential
is affected by this.

\subsection{Approach based on the linear response}\label{sec:tkp}

We here take the approach \cite{Thakur:2013nia} based on the linear response theory. 
The properties of a thermal medium is encoded 
in the dielectric permittivity $ \varepsilon(p)$. 
When the linear approximation is justified, 
the in-medium potential is related to the potential 
in the vacuum through the permittivity by 
\begin{equation}
V(\mathbf p) = V_{\rm vac}  (\mathbf p )  \varepsilon^{-1} (\mathbf p )  . 
\end{equation}
This relation is true even in a strongly coupled system, 
as long as the linear approximation to the potential is good. 
As a vacuum potential, we employ the Cornell potential. 
Thus, the in-medium heavy-quark potential in the real space 
can be written as 
\begin{eqnarray}
\label{defn}
V(r)&=&\int \frac{d^3\mathbf p}{{(2\pi)}^{3/2}}
(e^{i\mathbf{p} \cdot \mathbf{r}}-1)~\frac{V_{\rm Cornell} (p)}{\varepsilon(p)} ~,
\end{eqnarray}
where  the Fourier transform of the Cornell potential
$V_{\rm Cornell} (p) $ is given by 
\begin{equation}
V_{\rm Cornell} (p)=-\sqrt{(2/\pi)} \frac{\alpha}{p^2}-\frac{4\sigma}{\sqrt{2 \pi} p^4},
\label{eq:vcornell}
\end{equation}
where $ \alpha \equiv C_{F} \alpha_{s} $ with 
$C_{F}=(N_c^2-1)/2N_c$ and  $ \sigma $ is the string tension. 
The parameter $\sigma$ 
is determined to reproduce the vacuum quarkonium property.

\begin{figure}[tb]
	\subfigure{
		\hspace{-0mm}\includegraphics[width=7.7cm]{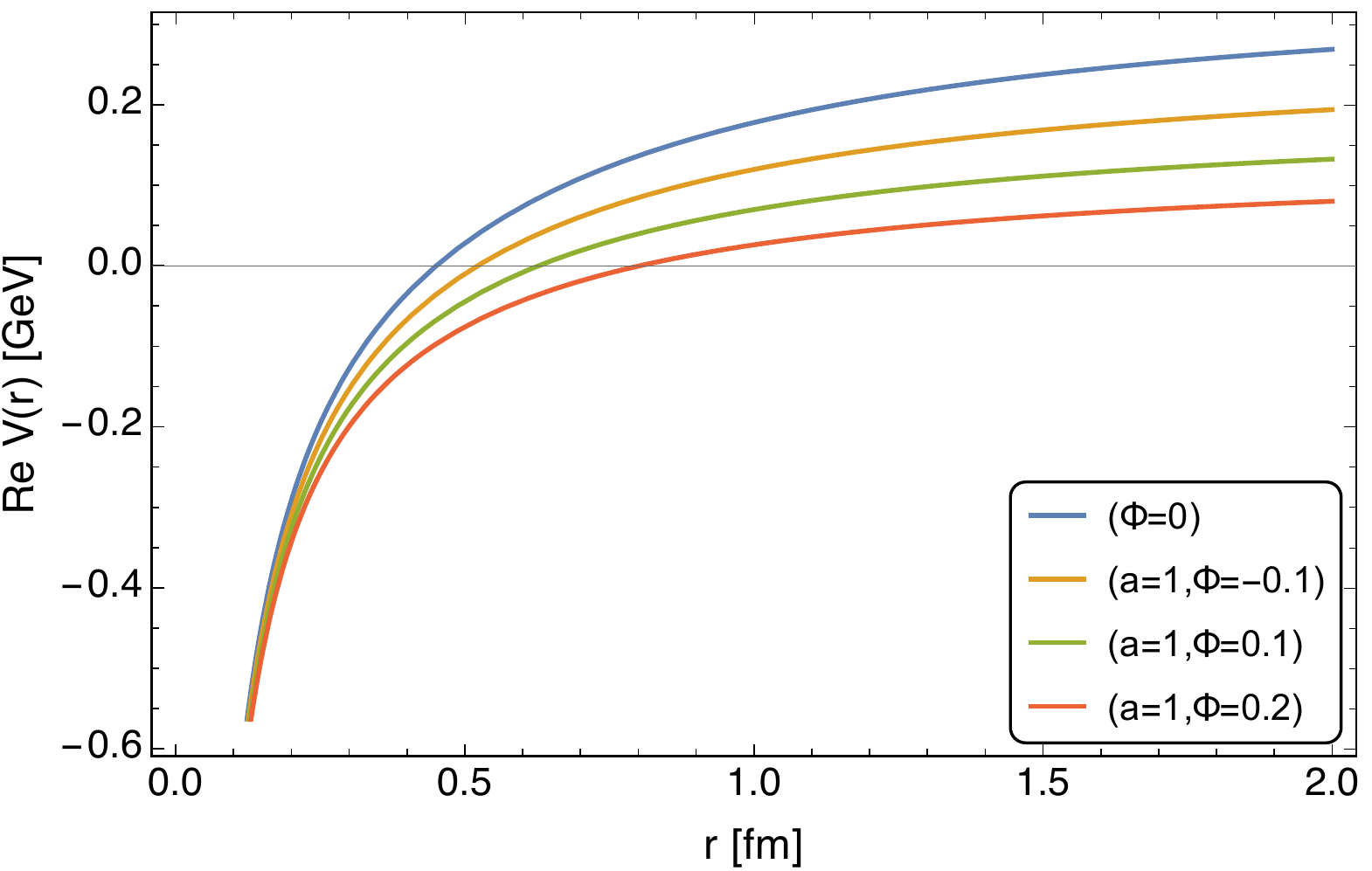}} 
	\subfigure{
		\includegraphics[width=7.7cm]{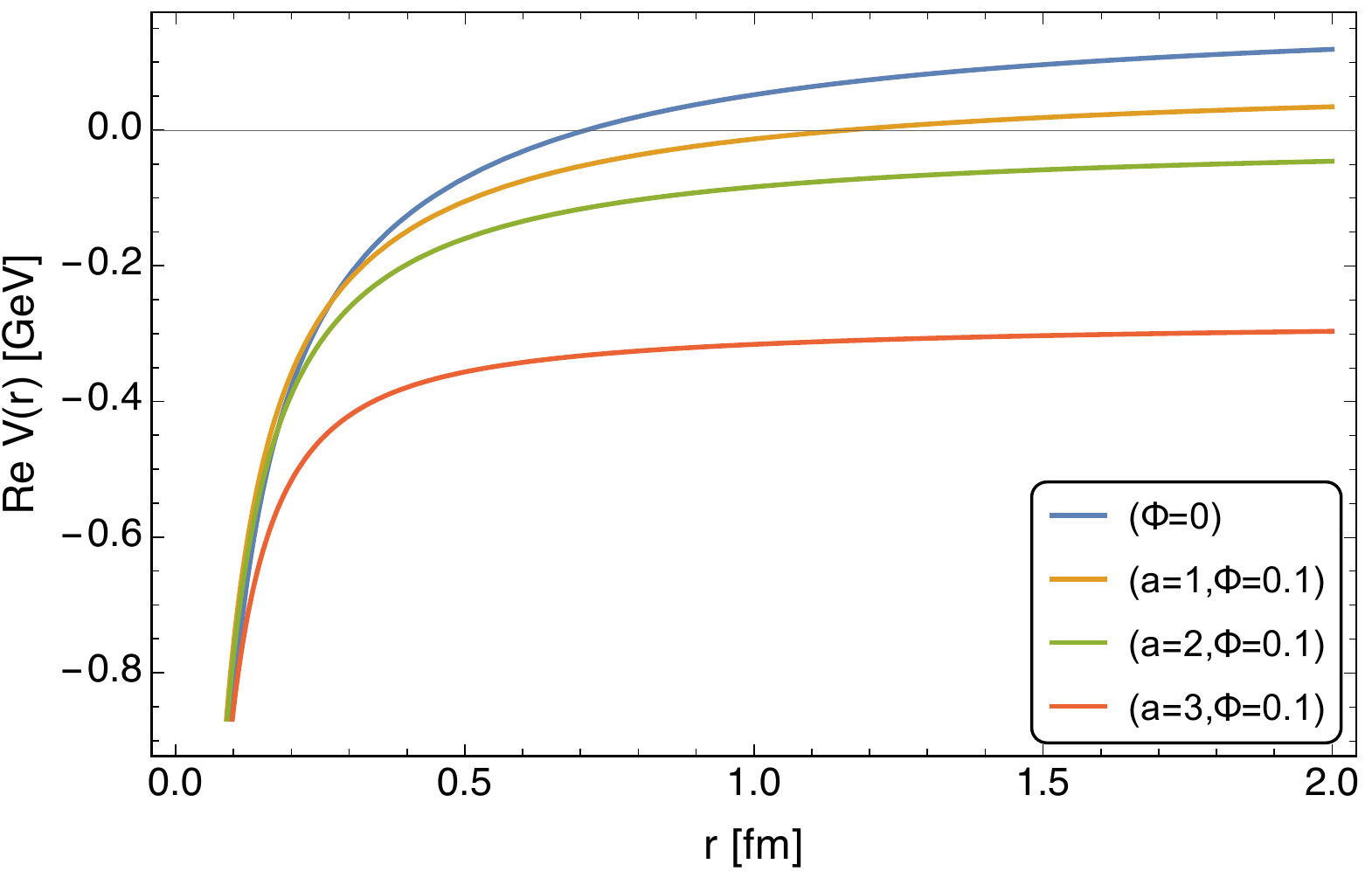}}
	\caption{
Left: real part of the potential ($ \Re V $) as a function of $r$
 for different values of $ \Phi $ with $ a=1 $ at $T= 0.3$ GeV. 
 Right: $ \Re V $  for different  values of $a$ with $\Phi= 0.1$ and
 $T=0.3$ GeV. 
The potential becomes more flattened  for larger values of $\Phi$ and $a$. 
}
	\label{pot_re_rmD}
\end{figure}

Let us first look at the real part of the in-medium heavy quark potential. 
Using Eqs.~(\ref{eq:epsilon}) and (\ref{eq:vcornell}), 
it is computed as 
\begin{equation}
\begin{split} 
\Re V(r)
&= 
\int \frac{d^3\mathbf p}{{(2\pi)}^{3/2}}
(e^{i\mathbf{p} \cdot \mathbf{r}}-1)
V_{\rm Cornell} (p)
\Re \varepsilon^{-1}(p)  \\
&= 
-\alpha \, 
\widetilde{m}_{D,R}
\left(\frac{e^{-\widetilde{m}_{D,R}\, r}}{\widetilde{m}_{D,R}\, r}+1\right) 
+ 
\frac{2\sigma}{
\widetilde{m}_{D,R} 
}\left(\frac{e^{-\widetilde{m}_{D,R}\,{r}}-1}{\widetilde{m}_{D,R}\,r}+1\right) .
\label{eq:pot-real}
\end{split}
\end{equation}
where the first term is Coulombic contribution with the Debye screening, 
and the second term comes from the string-like part of the Cornell potential. 
In the absence of bulk viscous corrections, this form of potential is derived 
in Ref.~\cite{Thakur:2013nia} . 
In the small distance limit, $r \to 0$, 
it approaches the Cornell potential, $V_{\rm Cornell}(r) =- \alpha / r + \sigma r$. 
For the real part of the potential, 
the bulk viscous correction enters 
through the modification of the Debye mass $m_D \to \widetilde{m}_{D, R}$. 
In Fig.~\ref{pot_re_rmD}, 
we plot the real part of the potential
for different values of $\Phi$ (left) and 
$a$ (right). The modified Debye mass is an increasing function 
of both $\Phi$ and $a$, and the potential becomes flattened 
for larger values of those parameters.

\begin{figure}[tb]
	\subfigure{
		\hspace{-0mm}
		\includegraphics[width=7.8cm]{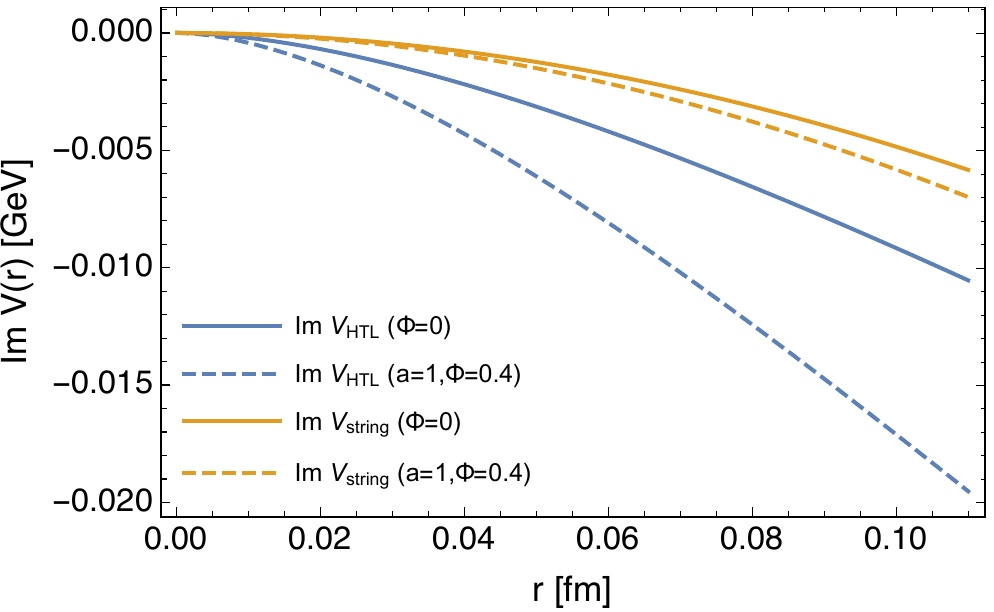}}
	\subfigure{
		\includegraphics[width=7.6cm]{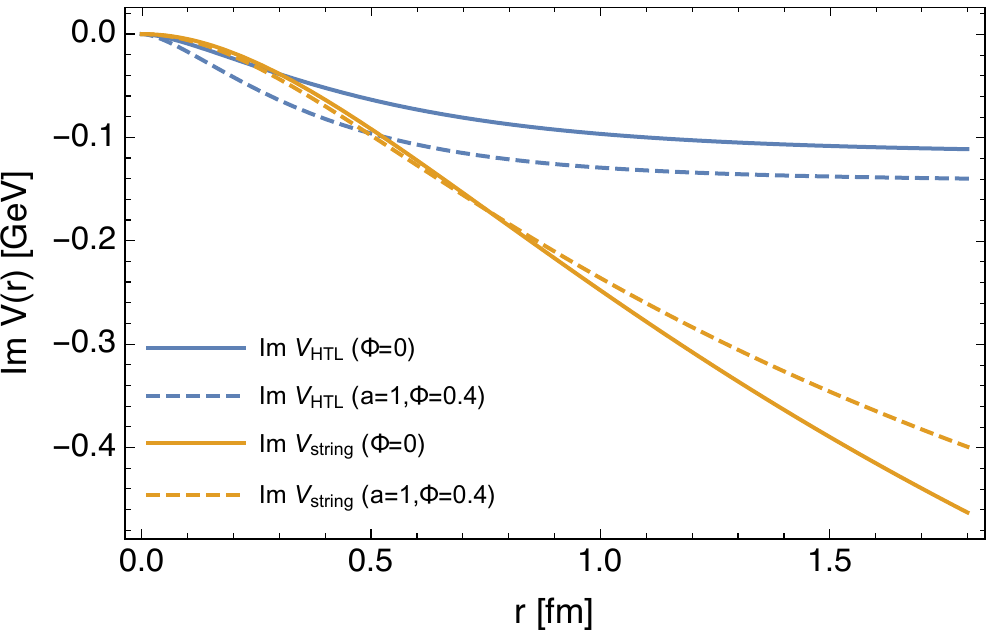}}
\caption{
Imaginary part of the potentials with and without bulk viscous corrections 
at $T=0.3 \,{\rm GeV}$.
The contribution from the HTL part  and the string-like part are shown separately. 
The left figure shows small $r$ region, 
and the right one shows a larger $r$ region. 
}
	\label{fig:imVSeparate}
\end{figure}

\begin{figure}[tb]
\centering
		\includegraphics[width=10cm]{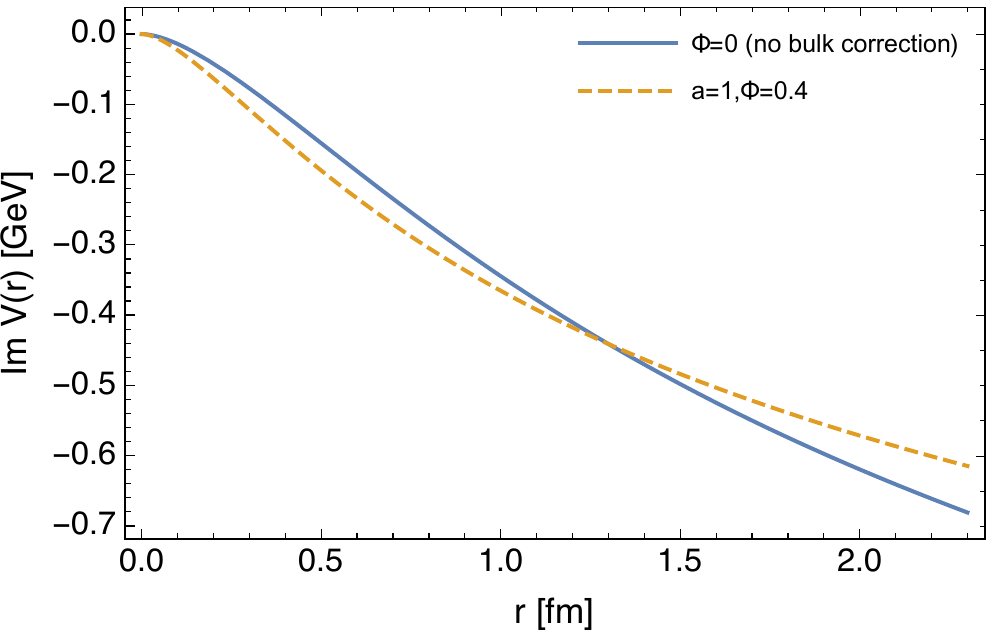}
\caption{
Imaginary part of the potential. 
$|\Im V|$ is enhanced in the small $r$ region, 
while it is suppressed at larger $r$. 
}
\label{fig:imVTotal}
\end{figure}

The heavy quark potential also acquires an imaginary part at finite temperatures.
The imaginary part reads 
\begin{equation}
\begin{split} 
\Im V(r)&=\int \frac{d^3\mathbf p}{{(2\pi)}^{3/2}}
(e^{i\mathbf{p} \cdot \mathbf{r}}-1)
V_{\rm Cornell}(p) \Im \varepsilon^{-1}  (p)  
\\
&= 
 -\alpha \lambda T 
\, \phi_2  (\widetilde{m}_{D,R}\, r)
- 
\frac{ 
2\sigma T 
\lambda 
 } {
  \widetilde{m}^2_{D, R} 
  }
 \, 
\chi( \widetilde{m}_{D, R}\, r ) \\
&\equiv
\Im V_{\rm HTL}(r)
+ 
\Im V_{\rm string}(r),
\label{eq:pot-imaginary}
\end{split}
\end{equation}
where the first term is from the perturbative HTL contribution, 
and the second term is the string-like contribution. 
We have plotted these terms separately in Fig.~\ref{fig:imVSeparate}, 
and the total imaginary part in Fig.~\ref{fig:imVTotal}. 
Note that the dimensionless parameter $\lambda$
is defined in Eq.~(\ref{eq:def-lambda}). 
The functions $\phi_n(x)$ and $\chi(x)$ are defined by 
\begin{equation}
\phi_n (x)
\equiv 
2 \int_{0}^{\infty} dz
 \frac{z}{(z^{2}+1)^{n}}\left[1-\frac{\sin(x z)}{x z}\right] ,
\end{equation}
\begin{equation}
\chi (x) 
\equiv
2 \int_{0}^{\infty}  \frac{dz}{z(z^{2}+1)^{2}}
 \left[1-\frac{\sin(x z)}{x z}\right]. 
\end{equation}
The function $\phi_2 (x)$ is a monotonically increasing function that asymptotes $\phi_2(0)=0$ and $\phi_2(\infty ) =1$. 
$\chi  (x)$ is also monotonically increasing 
with $\chi(0) =0$, 
but is logarithmically divergent at large $x$\footnote{
In Ref.~\cite{Lafferty:2019jpr}, the physical origin of the divergence is identified to the absence of string breaking, 
and a way  of regularization is discussed. 
On the value of the decay width that we later perform, 
this divergence is irrelevant because the wave function is localized. 
}.

Let us examine the qualitative features of the imaginary part and its 
bulk viscous correction. 
We plot the imaginary part of the potential in 
Fig.~\ref{fig:imVTotal}. 
As can be seen in the figure, the bulk viscous correction on $\Im V$ is different in small $r$ and large $r$ regions. 
This can be understood as follows: 
\begin{itemize}
\item
One consequence of the bulk viscous effect is a shift of the Debye mass, 
\begin{equation}
\frac{\widetilde m^2_{D, R}}{m^2_D}
= 1 + c_R(a) \Phi ,
\end{equation}
The Debye mass becomes heavier due to the bulk correction for $\Phi>0$
and $\phi( \widetilde{m}_{D, R} \, r)$ increases with the bulk correction. 
The coefficient of the Coulombic term is 
$\alpha \lambda$, which is also an increasing function of $\Phi$.  
So, $|\Im V_{\rm HTL}|$ is an increasing function of $\Phi$. 
The perturbative HTL contribution is dominant  in the small $r$ region, 
since $\phi_2 (\widetilde{m}_{D, R}\, r) \sim  r^2 \ln  r$ and 
$\chi \sim  r^2$ to the leading order. 
See the left panel of Fig. \ref{fig:imVSeparate}. 
Thus, in this region, $|\Im V|$ increases 
as $\Phi$ is increased. 
\item 
As shown in the right panel of Fig. \ref{fig:imVSeparate}, 
In the large $r$ region, string-like part dominates the imaginary part. 
In this region, $|\Im V_{\rm string}|$ is suppressed in the presence of 
bulk correction with $\Phi>0$. 
The string part is proportional to the factor 
$\lambda / \widetilde m^2_{D, R}$, 
which decreases as a function of $\Phi$. 
Although $\chi (\widetilde{m}_{D,R} \, r)$ is an increasing 
function of $\Phi$, in total the string part decreases as a function of $\Phi$.
\item As a result, for the bulk viscous correction for $\Phi>0$, 
we observe the enhancement of $|\Im V|$ in the small $r$ region, 
and  suppression at large $r$, 
as shown in Fig.~\ref{fig:imVTotal}.
\end{itemize}

\subsection{Introduction of non-perturbative propagator}

Let us discuss the prescription given in Ref.~\cite{Guo:2019bwa}. 
In order to take into account the string-like behavior of the potential, 
they have introduced a non-perturbative contribution to the
resummed gluon propagator in addition to the HTL contribution as
\begin{equation}
\bar D_R  =
 \bar D^{\rm p}_R  +  \bar D^{\rm np}_R , 
\label{eq:dp-dnp}
\end{equation}
where $\bar D^{\rm p}_R  =1 /(p^2 - \Pi_R  )$ is 
the perturbative HTL contribution. 
The non-perturbative contribution 
to the temporal component of the resummed gluon propagator 
is modeled by the following form, 
\begin{equation}
\bar D^{\rm np}_R (P)
= 
b 
\frac{ m_G^2 m_D^2 }
{( p^2 - \Pi_R  )^3
 }
+b' 
\frac{m_G^2 (- m^2_D - \Pi_{R} )} 
{( p^2 - \Pi_R )^3
 }, 
 \label{eq:gdpm-dnp}
\end{equation}
where $b=4$ and $b'=6$ is chosen
so that the leading contribution to the imaginary part of the potential 
in the small $r$ limit behaves as $r^4 \ln r$. 
The mass scale $m_G$ is related to the string tension as 
$m_G^2 = 2 \sigma / \alpha$ 
by matching the small $r$ behavior of the real part of the potential with the Cornell potential. 
To get the symmetric propagator,
the authors used the relation 
\begin{equation}
\bar D_S =  
\bar D_R \Pi_S \bar D_A 
=
\frac{\Pi_S}{
\Pi_R - \Pi_A
} 
\left( 
\bar D_R - \bar D_A
\right). 
\label{eq:ds-dr-da}
\end{equation}
Using the equilibrium self-energies (\ref{eq:pir-eq}) and  (\ref{eq:pis-eq})
in the HTL approximation, 
the non-perturbative part of the 
symmetric propagator in the static limit reads 
\begin{equation}
\bar D^{\rm np}_S(p_0 =0, \mathbf  p) =
12 \pi  i  T m^2_G m^2_D   \, 
\frac{p^2- m^2_G}
{
p (p^2 + m^2_D)^4
}. 
\label{eq:dsnp-eq}
\end{equation}
The Fourier transform of the propagators
leads to the potential\footnote{
The one given here 
is the second model discussed in Ref.~\cite{Guo:2019bwa}. 
In their first model, the non-perturbative retarded propagator is 
modeled as 
\begin{equation}
\bar D^{\rm np}_R (P) =  \frac{m_G^2}{ (p^2 - \Pi_R)^2 }. 
\end{equation}
The real part of this
propagator is $m_G^2 / (p^2+m_D^2)^2$ and 
 is the same as the non-perturbative 
propagator discussed in Ref.~\cite{Megias:2005ve}. 
This choice results in the KMS potential~\cite{Karsch:1987pv} for the real part. 
}
\footnote{
The same real part of the potential is obtained 
in Ref.~\cite{Strickland:2011aa} through a different line of reasoning. 
}
, 
\begin{eqnarray}
\Re V_{\rm GDPM} (r)
&=&
- \alpha 
\left( 
 \frac{e^{- m_D r}}{r}
 + m_D
  \right) 
+ 
 \frac{2 \sigma} {m_D}
(1 - e^{- m_D r})
-
\sigma r e^{- m_D r} , 
\label{eq:re-gdpm}
\\
\Im V_{\rm GDPM} (r)
&=& - \alpha \phi_2 (m_D  r) 
+ 
\frac{8 \sigma T}{m_D^2 }
\left[ 
 \phi_3 (m_D  r)
- 
3 \phi_4 (m_D  r)
\right] .
\end{eqnarray}

Let us make a comment on the derivation. 
The use of Eq.~(\ref{eq:ds-dr-da}) might look problematic, 
because when we modify the resummed propagator as  Eq.~(\ref{eq:dp-dnp}), 
the self-energies should in general be modified and be different from the perturbative one. 
The Dyson-Schwinger equation (\ref{eq:ds-s}) 
can be solved by the following expression, 
\begin{equation}
\bar D_S (P)
=
c \,{\rm sgn}(p_0)\, ( \bar D_R - \bar D_A )
-
c \,{\rm sgn}(p_0)\,
 \bar D_R (\Pi_R - \Pi_A ) \bar D_A 
 + \bar D_R \Pi_S \bar D_A . 
\label{eq:ds-3}
\end{equation}
The parameter $c$ can be in fact arbitrary, 
because 
the sum of the first two terms on the right hand side is zero. 
In the thermal equilibrium, the most convenient 
choice is  
$c = 1+2 f_0$, where $f_0 = f_0(p_0)$ is the Bose distribution. 
It is convenient because the FDT holds
in the thermal equilibrium, 
\begin{equation}
\Pi_S (P)= (1+2 f_0) \,{\rm sgn}(p_0)\, ( \Pi_R - \Pi_A ), 
\end{equation}
because of which the last two terms 
of Eq.~(\ref{eq:ds-3}) are equal in magnitude. 
Therefore, in equilibrium, the symmetric propagator 
can be expressed in three equivalent ways, 
\begin{equation}
\begin{split} 
\bar D_S (P) &=
(1+2 f_0) \,{\rm sgn}(p_0)\, (\bar D_R - \bar D_A )  \\
&= (1+2 f_0) \,{\rm sgn}(p_0) \bar D_R (\Pi_R - \Pi_A) \bar D_A  \\
&= \bar D_R \Pi_S \bar D_A . 
\end{split}
\end{equation}
If we use the first expression, 
it is evident that when 
$\bar D_R$ has an additive contribution as
Eq.~(\ref{eq:dp-dnp}), 
the symmetric propagator gets an additive contribution as 
$\bar D_S = \bar D^{\rm p}_{S} + \bar D^{\rm np}_{S} $. 
Namely, Eq.~(\ref{eq:ds-dr-da})  does not actually depend 
on the self-energies. 

From this consideration, it might seem difficult 
to extend this prescription to non-equilibrium. 
This is because, 
we do not have the FDT in non-equilibrium,
and we have to use Eq.~(\ref{eq:ds-dr-da}) to compute the resummed symmetric propagator, 
but there is no way to determine $\Pi_S$ when there are perturbative and non-perturbative contributions in $\bar D_R$. 
However, in the current case of the bulk viscous correction, 
we can determine the $\bar D_S$ even though it is non-equilibrium, 
by assuming 
a modified FDT  (\ref{eq:mfdt-pi})\footnote{
The modified FDT is derived in the perturbation theory 
and whether it is still valid in the non-perturbative regime can be questioned. 
}.
We can choose the parameter as $c = 2 T \lambda / p_0$
and we can express $\bar D_S$ as 
\begin{equation}
\bar D_S (P)
=
\frac{2 T \lambda}{p_0}   {\rm sgn}(p_0) \,( \bar D_R - \bar D_A )
-
\frac{2 T \lambda}{p_0}  {\rm sgn}(p_0) \,
 \bar D_R (\Pi_R - \Pi_A ) \bar D_A 
 + \bar D_R \Pi_S \bar D_A . 
\label{eq:ds-3-noneq}
\end{equation}
Because of the modified FDT (\ref{eq:mfdt-pi}), 
and the last two terms in Eq.~(\ref{eq:ds-3-noneq}) cancel, 
just like the case of equilibrium. 
Therefore, we can express the resummed symmetric propagator 
in three ways, 
\begin{equation}
\begin{split} 
\bar D_S (P) &=
 \frac{2 T \lambda}{p_0} {\rm sgn}(p_0) ( \bar D_R - \bar D_A ) 
 \\
&= \frac{2 T \lambda}{p_0} \,{\rm sgn}(p_0) \bar D_R (\Pi_R - \Pi_A) \bar D_A  \\
&= \bar D_R \Pi_S \bar D_A . 
\end{split}
\label{eq:fdt-ds-noneq}
\end{equation}
In the first expression, it is evident that
when the retarded (or advanced) propagator gets 
an additive contribution as Eq.~(\ref{eq:dp-dnp}), 
the non-perturbative contribution 
additively contribute to $\bar D_S = \bar D^{\rm p}_S + \bar D^{\rm np}_S $. 

In the presence of bulk viscous corrections, 
let us we consider the same form of the non-perturbative contribution to the retarded 
self-energy, 
\begin{equation}
\bar D^{\rm np}_R  (P)
= 
b 
\frac{ m_G^2 \widetilde m^2_{D,R} }
{( p^2 - \Pi_R  )^3
 }
+b' 
\frac{m_G^2 (- \widetilde m^2_{D,R} - \Pi_{R} )} 
{( p^2 - \Pi_R )^3
 }, 
 \label{eq:gdpm-dnp-bulk}
\end{equation}
where $\Pi_R$ is given by Eq.~(\ref{eq:pir-bulk}). 
Note that the second term does not contribute to the 
real part of the potential, because its real part vanishes in the static limit 
$p_0 \to 0$. 
We can compute the corresponding 
non-perturbative part of the resummed symmetric propagator using Eq.~(\ref{eq:fdt-ds-noneq}) as 
\begin{equation}
\bar D^{\rm np}_S(p_0 =0, \mathbf  p) =
12 \pi  i  T  \lambda \,
m^2_G \widetilde m^2_{D,R}   \, 
\frac{p^2- m^2_G}
{
p (p^2 + \widetilde m^2_{D,R})^4
}. 
\label{eq:dsnp-bulk}
\end{equation}
The difference from the equilibrium case (\ref{eq:dsnp-eq}) is 
the replacement $m_D \to \widetilde m_{D, R}$ and 
the multiplication by $\lambda$.

\begin{figure}[tb]
	\subfigure{
		\hspace{-7mm}
		\includegraphics[width=7.7cm]{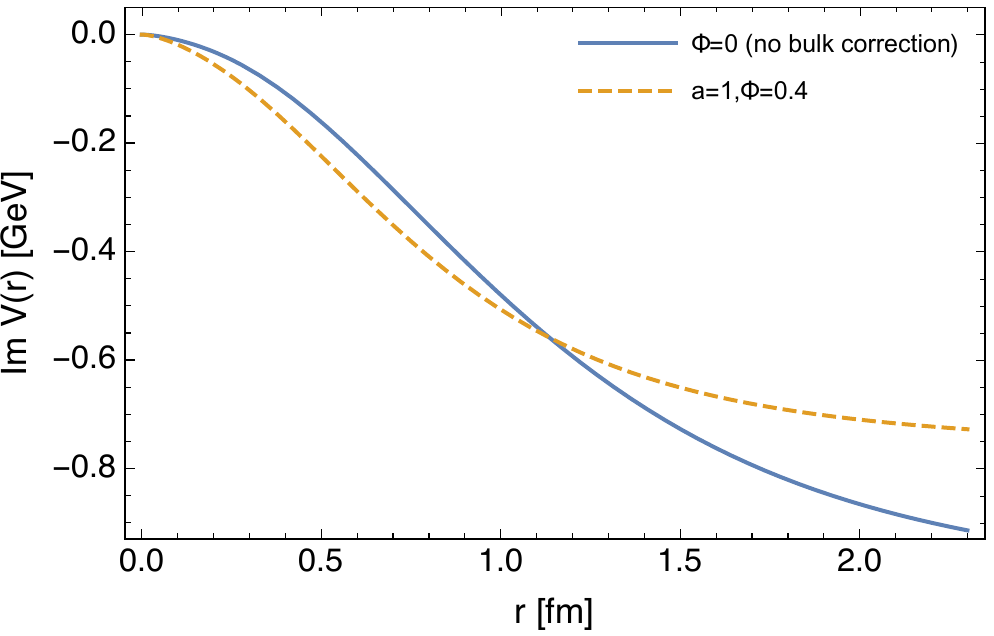}}
	\subfigure{
		\includegraphics[width=7.7cm]{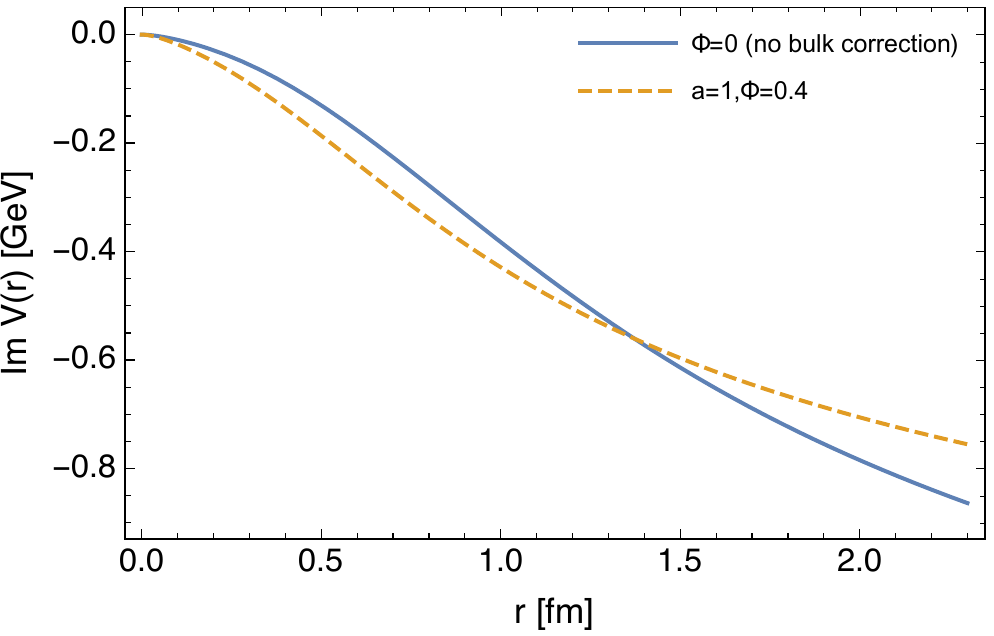}}
\caption{
Imaginary part of the potential
obtained via the prescription of 
Ref.~\cite{Guo:2019bwa} (left) and Ref.~\cite{Lafferty:2019jpr} (right). 
The potentials without bulk viscous correction are denoted by solid lines, 
and dashed lines are with the corrections. 
In both prescriptions, 
$|\Im V|$ is enhanced in the small $r$ region, 
while it is suppressed at larger $r$. 
The parameters are set as $T=0.3 \, {\rm GeV}$, $\sigma = (0.44)^2\, {\rm GeV^2}$, and $\alpha = 0.392$. 
}
\label{fig:imVOther}
\end{figure}

Now that we got the expression of the necessary 
propagators and we are ready to compute the potential 
through Eq.~(\ref{eq:d11}). 
The real part can be obtained by a simple replacement 
of the Debye mass $m_D \to \widetilde m_{D, R}$. 
Using the modified symmetric propagator (\ref{eq:dsnp-bulk}), 
the imaginary part of the potential is given by 
\begin{equation}
\Im V_{\rm GDPM}^{\rm bulk }(r)
=
 - \alpha \lambda  \phi_2 (\widetilde{m}_{D,R} \, r) 
+ 
\frac{ 8 \sigma T  \lambda  }{ \widetilde m^2_{D, R}}
\left[  
  \phi_3 (\widetilde{m}_{D,R} \, r)
- 3  \phi_4 (\widetilde{m}_{D,R}\,  r)
\right] . 
\end{equation}
On the left panel of Fig.~\ref{fig:imVOther}, we plot the imaginary part of the potential for both equilibrium and non-equilibrium cases. 
Qualitatively, how the imaginary part is affected is the same as the case 
(\ref{eq:pot-imaginary}). 
$|\Im V|$  gets enhanced at small $r$, 
and suppressed at large $r$.

\subsection{Approach based on a generalized Gauss law}

In Ref.~\cite{Lafferty:2019jpr}, a different prescription to 
obtain the potential is given. 
They have used a generalized Gauss law, 
that can be applicable to a string-like potential, 
and the HTL permittivity to obtain an analytic form of the in-medium potential. 
The real part of the potential is the same as Eq.~(\ref{eq:re-gdpm}). 
The  imaginary part of the potential in the thermal equilibrium is 
given by 
\begin{equation}
\Im V_{\rm LR} (r)=
- \alpha T \phi_2(m_D r) 
- 
\frac{ \sqrt{\pi}}{4}
T \sigma 
m_D 
r^3 \, G_{2,4}^{2,2}
\left( 
\begin{matrix}
-\frac{1}2, -\frac{1}2 \\
\frac{1}2,\frac{1}2, -\frac{3}2,-1 \\
\end{matrix}
\middle|
\frac{m_D^2 r^2}{4}
\right), 
\end{equation}
where $G$ is Meijer’s $G$-function. 
We can use the same prescription 
straightforwardly to obtain the 
potential in the presence of the bulk viscous corrections, 
by using the modified dielectric permittivity (\ref{eq:epsilon}). 
The real part is again just a replacement $m_D \to \widetilde{m}_{D,R}$. 
The imaginary part is given by 
\begin{equation}
\Im V^{\rm bulk}_{\rm LR} (r)= 
- \alpha \lambda T \phi_2(\widetilde{m}_{D,R} \, r) 
- 
\frac{ \sqrt{\pi}}{4}
 T \sigma  \lambda\, \widetilde m_{D, R}\, r^3 
 \, G_{2,4}^{2,2}
\left( 
\begin{matrix}
-\frac{1}2, -\frac{1}2 \\
\frac{1}2,\frac{1}2, -\frac{3}2,-1 \\
\end{matrix}
\middle|
\frac{ \widetilde{m}_{D,R}^2\, r^2}{4}
\right) . 
\label{eq:lr-imag}
\end{equation}
In this prescription too, the way the bulk correction affects 
the potential is also very similar to the case (\ref{eq:pot-imaginary}). 
At small $r$, the second term is $r^4$ while the first term is $ r^2 \ln  r$, so the first term is dominant, where its magnitude is enhanced for $\Phi>0$. 
At large $r$, the second term is dominant. 
On the right panel of Fig.~\ref{fig:imVOther}, 
we plot the imaginary part (\ref{eq:lr-imag}) with and without bulk corrections.
For $\Phi>0$, 
$|\Im V_{\rm LR}|$ is enhanced in the small $r$ region, 
and suppressed at large $r$. 

\begin{figure}[tb]
\centering
		\includegraphics[width=13.5cm]{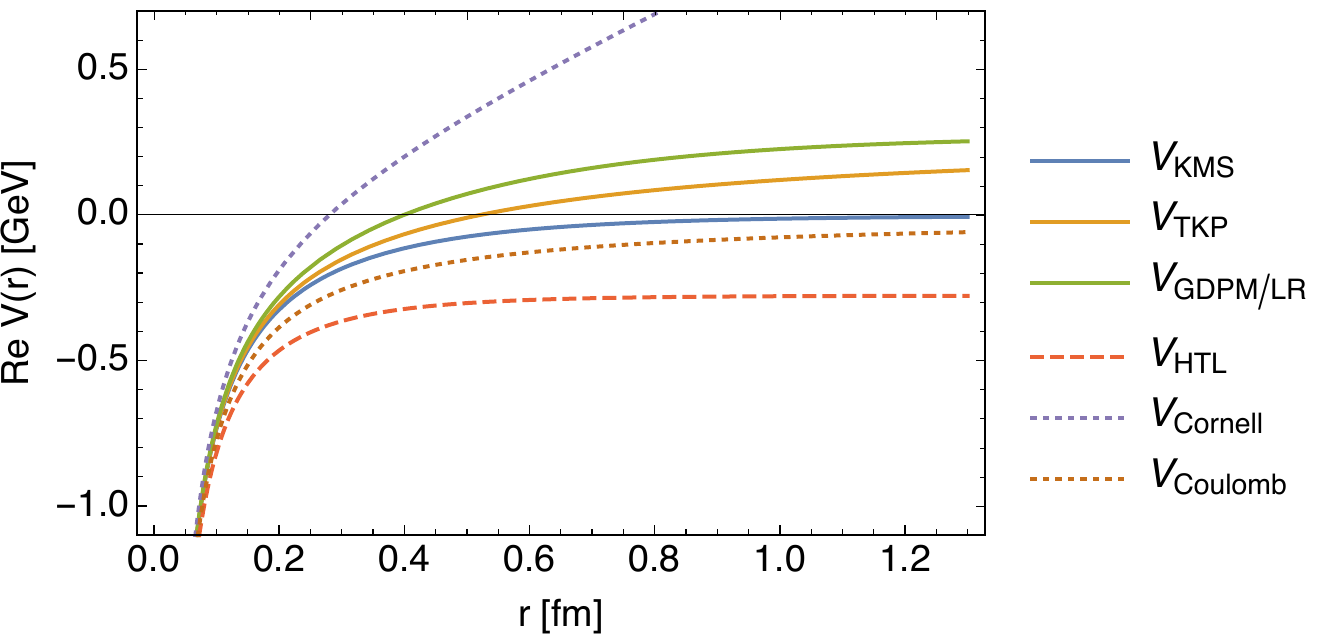}
\caption{
Comparison of the models of the real part of the in-medium potential. 
}
\label{fig:realPotentials}
\end{figure}

\subsection{ Comparison of the real part of the potentials }

Let us conclude this section by giving a comparison of 
several models of heavy quark potentials in the thermal equilibrium. 
In Fig.~\ref{fig:realPotentials}, we plot the potentials 
 (\ref{eq:pot-real}) (denoted by $V_{\rm TKP}$)
 and (\ref{eq:re-gdpm}) (denoted by $V_{\rm GDPM/LR}$) 
in the absence of bulk viscous corrections. 
We have also plotted one of the previous model of the finite temperature potential, given by 
\cite{Karsch:1987pv} 
\begin{equation}
V_{\rm KMS} (r)
=
- \frac{\alpha}{r} e^{- m_D r}
+ \frac{\sigma} {m_D}
(1 - e^{- m_D r}). 
\end{equation}
All the models are parametrized by the Debye mass $m_D$, 
strong coupling constant $\alpha$, and the string tension $\sigma$. 
We take common values for the plot. 
Namely, all the models asymptote to the 
Cornell potential with the same parameters  in the small distance limit. 
For a comparison, we also plotted 
the perturbative HTL potential (the first term of Eq.~(\ref{eq:pot-real})), 
the Cornell potential, and Coulomb potential. 
The potentials $\Re V_{\rm KMS}, \Re V_{\rm TKP}$, and  $\Re V_{\rm GDPM/LR}$ start to deviate from each other around the 
length scale given by the inverse Debye mass.

\section{
Heavy quarkonia in the presence of bulk viscous corrections
}\label{sec:hq-property}

In this section, we discuss the effect of bulk corrections 
on the properties of heavy quarkonia, 
such as the binding energies and decay widths, 
based on the potential obtained in the section~\ref{sec:tkp}. 

\subsection{Computational setup}

Let us describe the computational procedure. 
In order to study the in-medium properties 
of quarkonia, 
we here solve the Schr\"odinger equation for a heavy quarkonium
to obtain the wave function, using the real part of the in-medium potential.

The potential is the function of only the radial coordinate 
and we only have to solve the ordinary differential 
equation of the radial part of the wave function. 
The time-independent 
Schr\"{o}dinger equation for the radial wave function reads 
\begin{equation}
- \frac{1}{2 m_q}
\left( 
\psi'' (r) + \frac{2}r \psi' (r)
- \frac{ \ell (\ell+1)}{r^2} \psi (r)
\right)
+ 
 \Re V(r) \,  \psi  (r)
 = 
\epsilon_{_{n\ell}} \,  \psi(r) ,
\end{equation}
where $m_q$ is the reduced mass of the quarkonium system. 
We numerically solve this using the real part of the potential (\ref{eq:pot-real})
modified by the bulk viscous correction
to obtain the wave functions and eigenvalues\footnote{
	The interpretation of the role of  a complex potential 
	needs some care. 
	As shown in \cite{Burnier:2007qm}, 
	the complex potential dictates the time evolution of 
	unequal time point-split meson-meson correlator, 
	which is related to spectral functions, and 
	the binding energies and 
	decay widths can be read off from the spectral functions. 
	Strictly speaking, the potential does not describe the time evolution of the  wave function itself. 
	In this study, we used the real part to dictate the wave function itself
	and computed the binding energies and decay widths based on the 
	wave function. 
	An advantage of this approach is that it gives us an intuitive picture
	on the behavior of heavy quarkonia in the medium. 
	The resultant melting temperature is in agreement with 
	those obtained from the approach \cite{Burnier:2007qm}. 
	When the decay width becomes comparable to the binding energy, 
	it would be better to directly compute the spectral function. 
}.

The binding energy is given by  the 
difference between the asymptotic value of the potential
and the eigenvalue $\epsilon_{_{n\ell}}$, 
\begin{equation}
E_{\rm bin} = \Re V(r \to \infty) - \epsilon_{_{n\ell}} ,
\end{equation}
In the case of the potential (\ref{eq:pot-real}), the asymptotic value is 
\begin{equation}
\Re V (r \to \infty ) 
= - \alpha \widetilde{m}_{D,R} + \frac{2\sigma}{\widetilde{m}_{D,R}} . 
\end{equation}

Using the imaginary part of the potential (\ref{eq:pot-imaginary}), 
we can make an estimate for the thermal decay  width $\Gamma$ by 
\begin{equation}
\label{eq:thermalwidth}
\Gamma
= - \langle \psi | \Im \, V(r) | \psi \rangle 
=
- 
\frac{
 \int dr\, r^2 |\psi(r)|^2  \Im V(r)
}{
 \int dr \, r^2 |\psi(r)|^2 
} . 
\end{equation}
In a thermal environment, the wave function cannot exist as a steady state, 
but is transient. 
This way of the estimate of the decay width treat the imaginary part of the potential as a perturbation and its validity worsens when the decay width become comparable the binding energy.

Below is how the parameters are set: 
\begin{itemize}
\item 
For the Debye masses, 
we use the perturbative expressions 
(\ref{mDRtot}), (\ref{mDStot}) for different temperatures/chemical potentials, including the bulk viscous corrections. 
Since potentials are parametrized by  the Debye mass
one could regard the Debye masses as a parameter and 
fit it to reproduce the potential computed from the lattice QCD data. 
We do not take this approach here and use the HTL expression, 
since the primary  focus in this study is to understand 
the nature of bulk viscous modification on heavy-quark properties. 
\item 
We set the reduced heavy quark mass to 
$m_q = 1.25 /2 \,{\rm GeV} $ for $c \bar c$ 
and 
$m_q = 4.66 /2 \, {\rm GeV} $ for $b \bar b$.  
\item 
The string tension is chosen to  $ \sigma=(0.44 \, {\rm GeV})^{2}$. 
\item 
For the coupling constant, 
we use the one-loop result, 
\begin{equation}
\alpha_s 
=
\frac{g^2}{4 \pi }
= \frac{6 \pi}{
(11 N_c - 2 N_f ) \log \left(
 2 \pi \sqrt{T^2+\mu^2/\pi^2 }
 / \Lambda \right)
}, 
\end{equation}
with $N_c=N_f=3$. 
Namely the renormalization scale is taken to be $2 \pi \sqrt{T^2+\mu^2/\pi^2 }$. 
The scale $\Lambda$ is chosen to $\Lambda=0.176\,{\rm GeV}$
requiring $\alpha_S (1.5 \, {\rm GeV})=0.326$ is satisfied 
to match the lattice measurements \cite{Bazavov:2012ka}. 
\end{itemize}
Under the current setting, we expect out computation becomes 
less trustable at lower temperatures. The Debye mass would deviate 
from $\propto T$ behavior, and the value of the coupling would deviate from the 1-loop results. This point may be improved by extracting the Debye mass and the coupling from lattice QCD results.

\subsection{Wave function}

\begin{figure}[tb]
	\subfigure{
		\hspace{-6mm}
		\includegraphics[width=7.6cm]{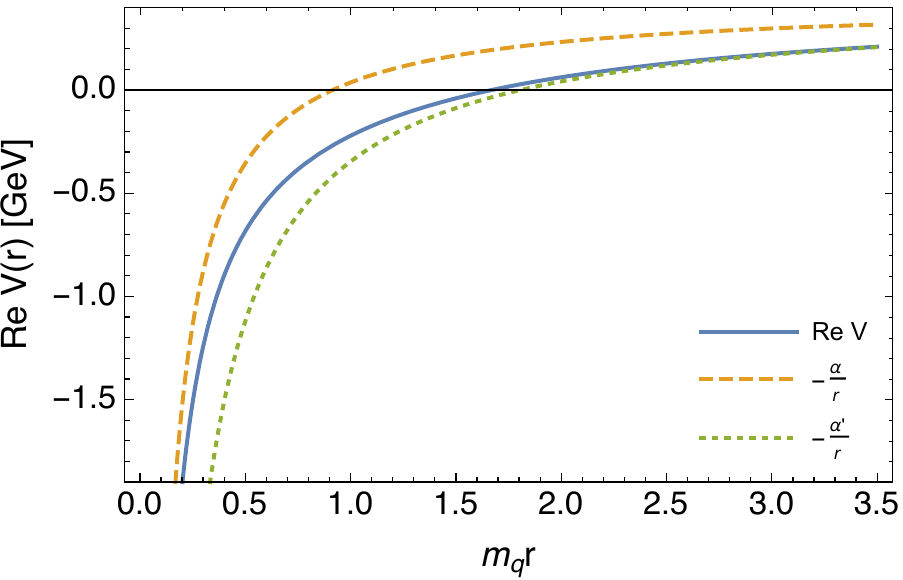}}
	\subfigure{
		\includegraphics[width=7.9cm]{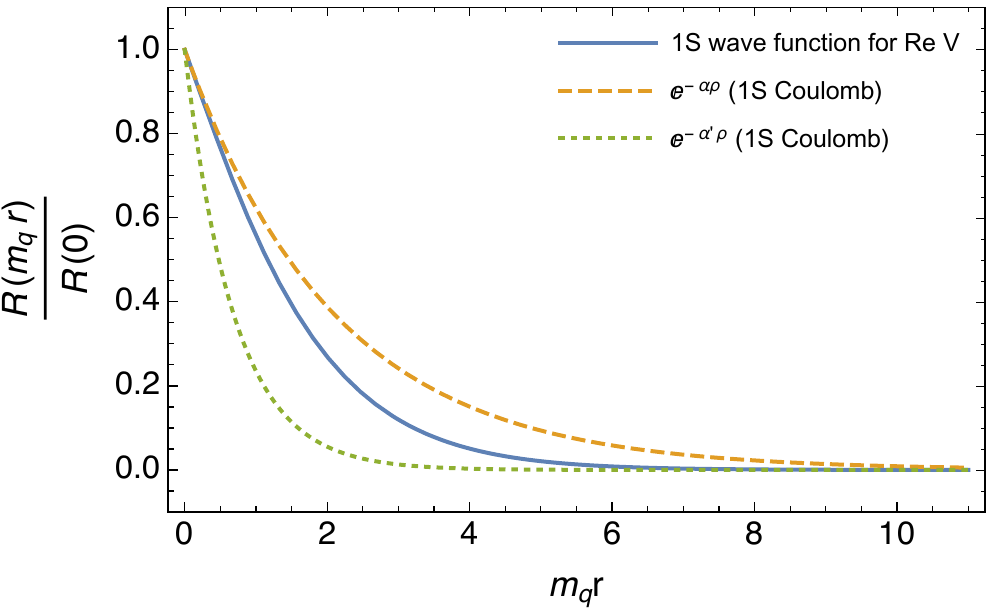}}
\caption{
Left: real part of the in-medium potential (\ref{eq:pot-real})  
at $T=0.3\,{\rm GeV}$, $\mu=0$ and $\Phi=0$ 
in comparison with its asymptotic Coulomb potentials. 
Right: radial wave function $R(m_q r)$ of $J/\psi$ state. 
The ground states of the asymptotic Coulomb potentials are shown for comparison. 
}
	\label{fig:potential-wf}
\end{figure}

Let us first discuss the qualitative feature of the wave functions 
of quarkonia from the finite temperature potential (\ref{eq:pot-real}). 
At $r \to 0$, 
the potential approaches a Coulomb potential
with coefficient $\alpha$, 
\begin{equation}
\Re V \sim - \frac{\alpha}{ r } . 
\end{equation}
On the  other hand, at large distances, 
the potential also approaches a  Coulomb potential,
but with a  different coefficient, 
\begin{equation}
\Re V \sim 
-  \frac{ \alpha' }{r}, 
\end{equation}
with $\alpha' \equiv 2 \sigma / \widetilde{m}^2_{D, R}$. 
The switching of those two regimes happens around 
$r$ given by the inverse Debye mass. 
In the limit of the large Debye  mass (or high temperature),
this part is flattened since the coefficient $\alpha'$ goes to zero, 
and the potential is dominated by the screened HTL one. 
In the left panel of Fig.~\ref{fig:potential-wf}, 
we plot the potential as well as its asymptotic Coulomb potentials. 
In the right panel of Fig.~\ref{fig:potential-wf}, 
we show the ground state wave function 
from the potential (\ref{eq:pot-real}) normalized at $r=0$ , 
as well as those from the asymptotic Coulomb potentials, 
at $T=0.3\,{\rm GeV}$. 
At this temperature, the Debye mass $\sim 0.63 \, {\rm GeV}$ is comparable to the reduced mass of the charm quark, and 
the wave function is away from both of the two Coulomb wave functions. 

\begin{figure}[tb]
	\subfigure{
		\hspace{-7mm}
		\includegraphics[width=7.7cm]{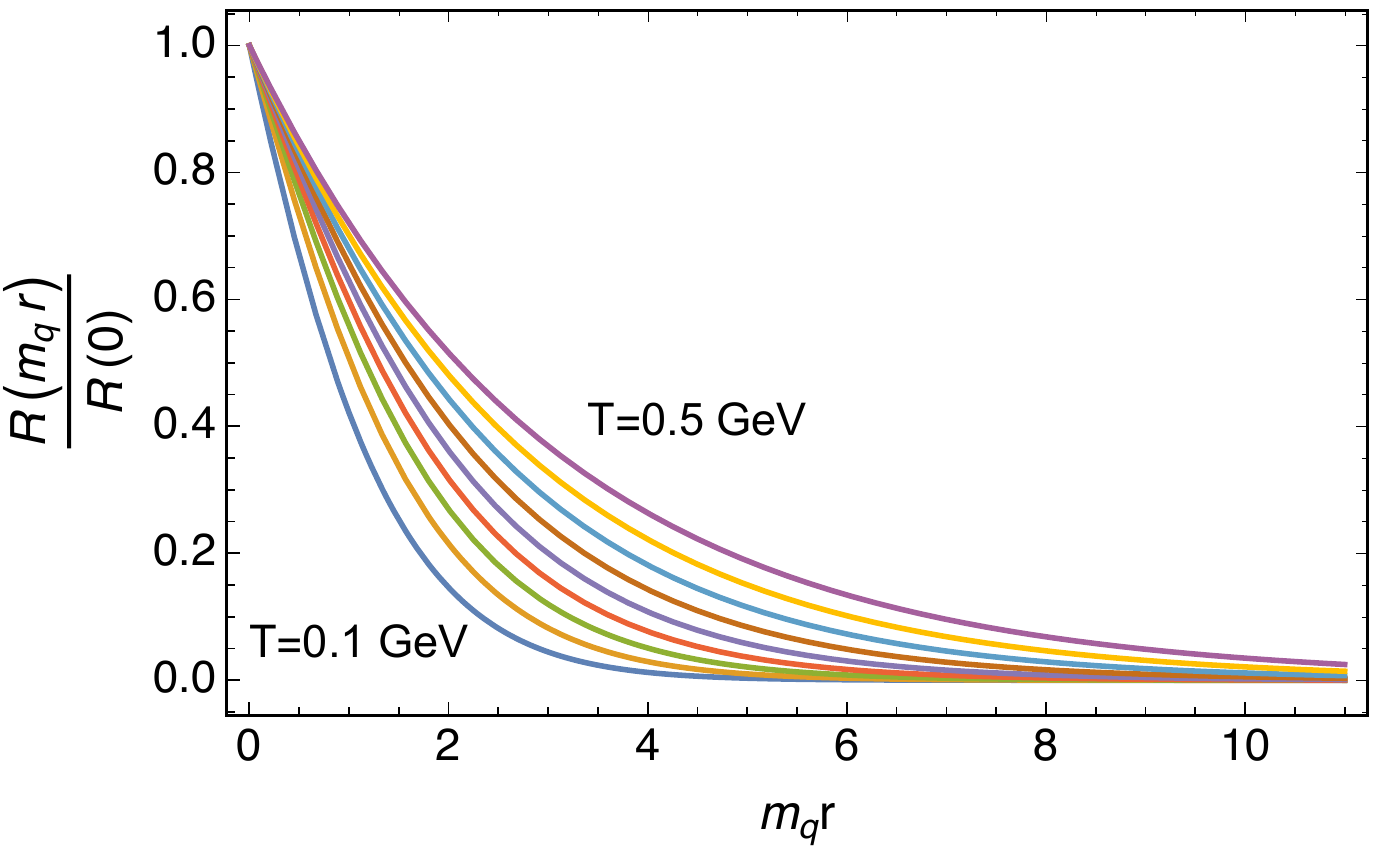}}
	\subfigure{
		\includegraphics[width=7.8cm]{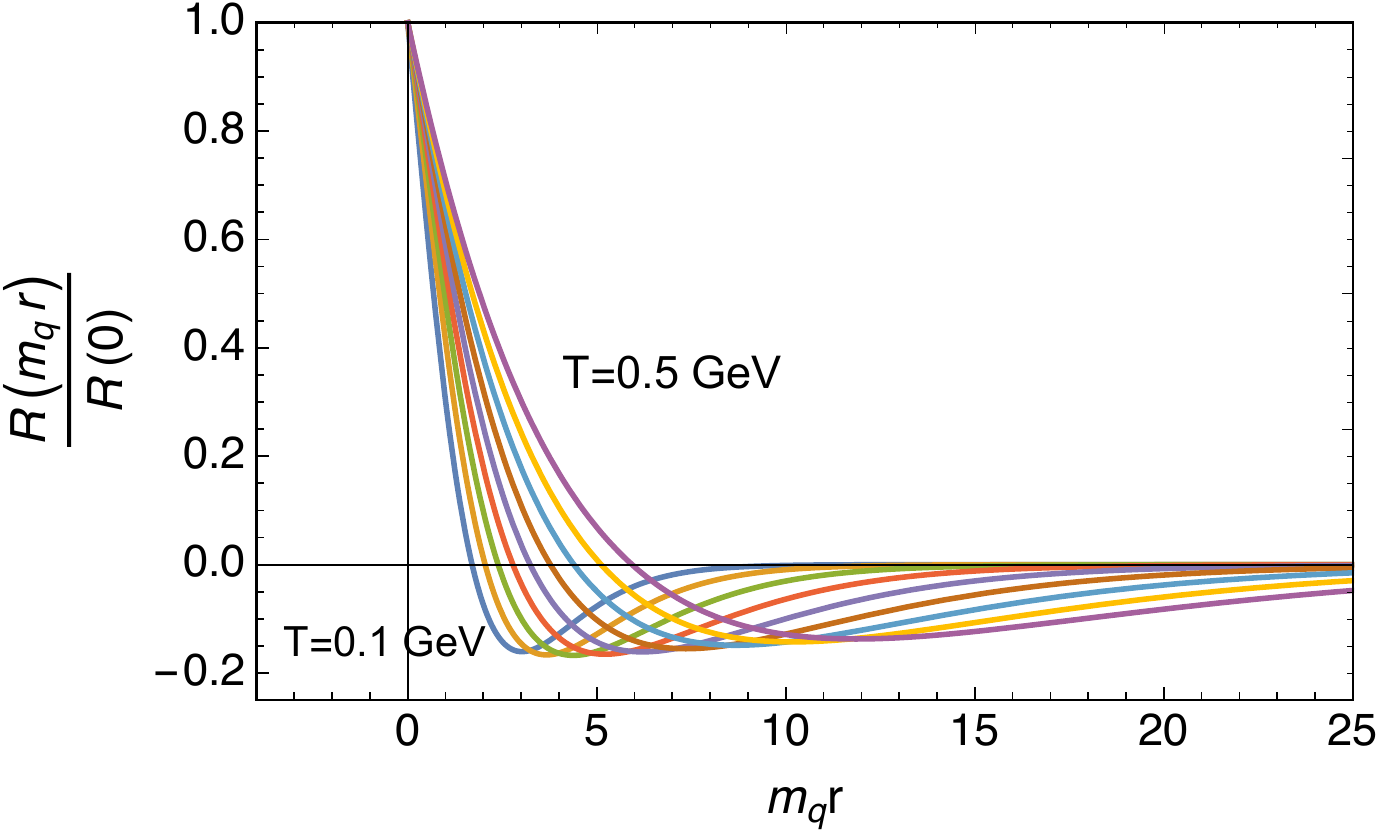}}
\caption{
Radial wave functions of the ground state (left) and first excited state (right) 
of $c \bar c$ wave functions at different temperatures, 
as a function of $m_q r$. 
The temperature is changed from $0.1\, {\rm GeV}$ 
to $0.5\, {\rm GeV}$ by $0.05 \,{\rm GeV}$. 
}
	\label{fig:wf-T}
\end{figure}
\begin{figure}[tb]
	\begin{minipage}[b]{0.5\linewidth}
	  \centering
	  \includegraphics[keepaspectratio, scale=0.55]{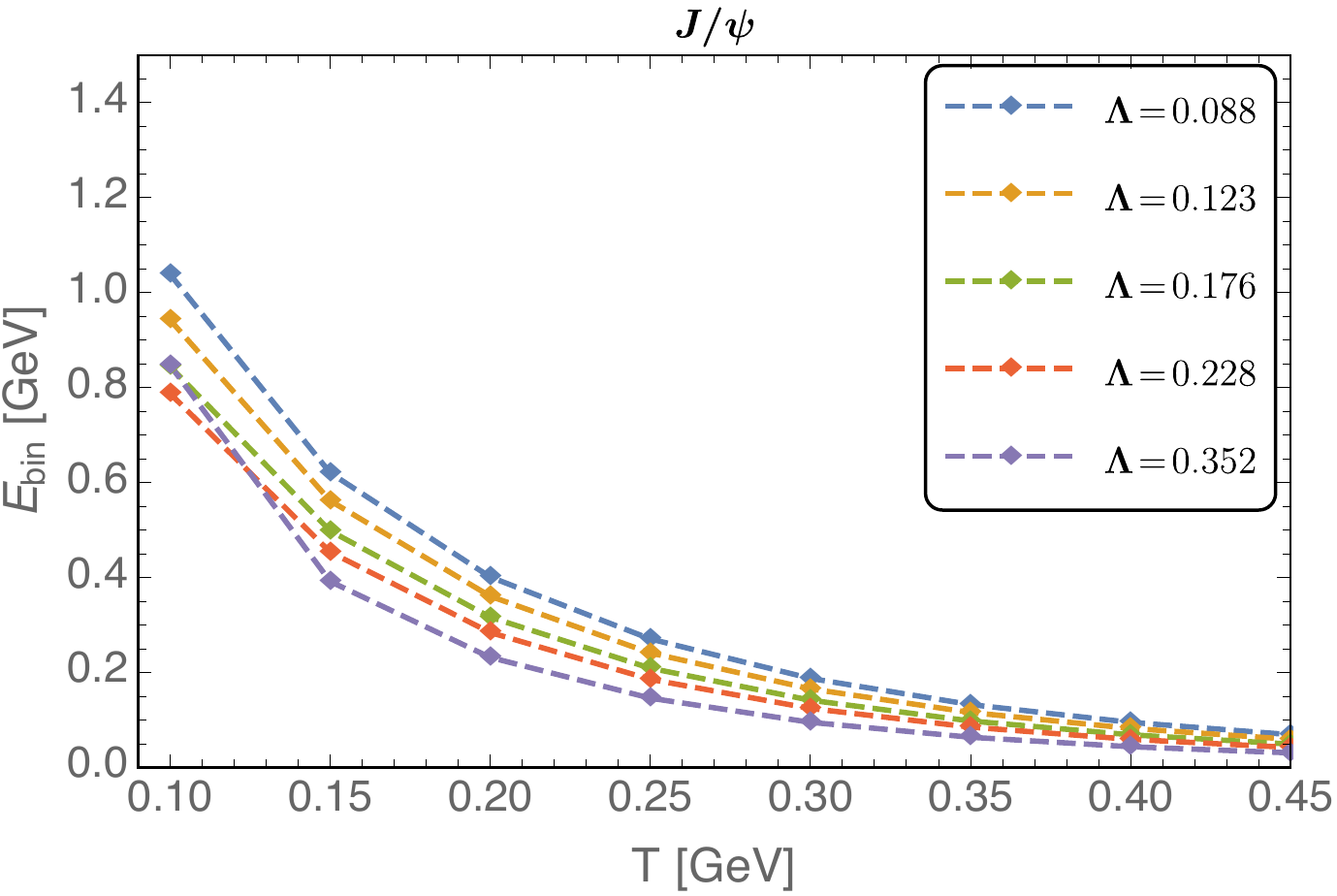}
	\end{minipage}
	\begin{minipage}[b]{0.5\linewidth}
	  \centering
	  \includegraphics[keepaspectratio, scale=0.55]{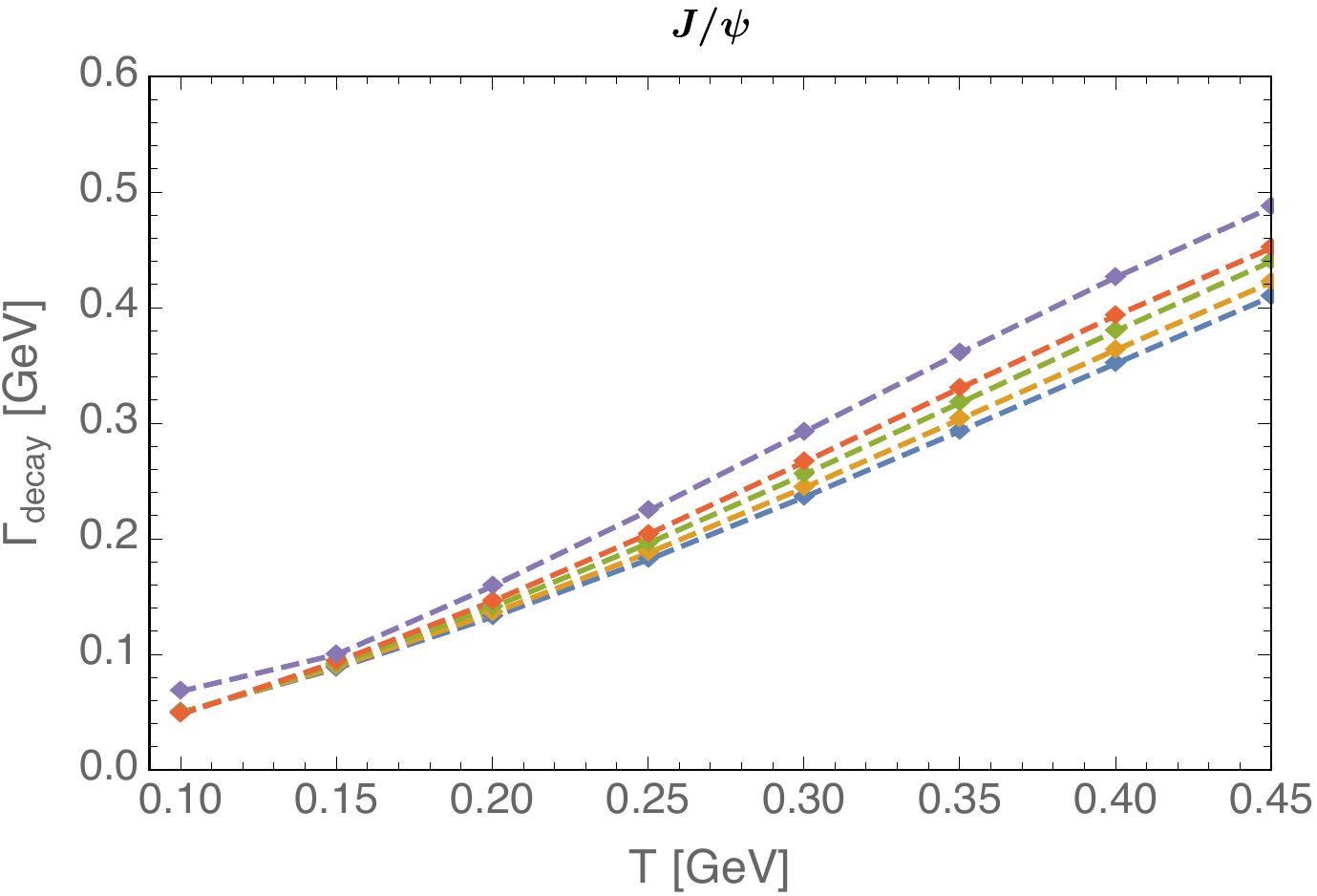}
	\end{minipage} \\
	\begin{minipage}[b]{0.5\linewidth}
		\centering
		\includegraphics[keepaspectratio, scale=0.55]{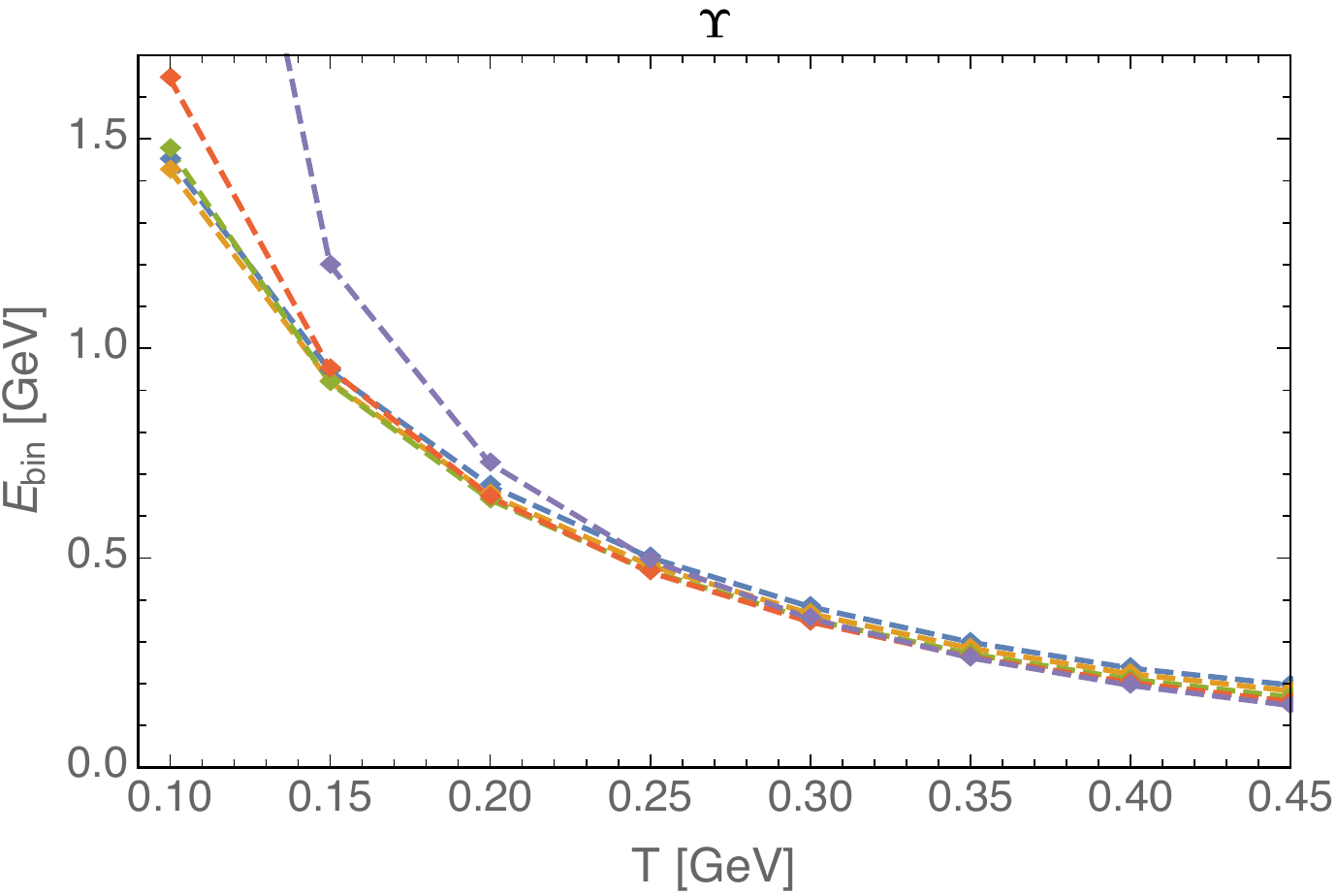}
	  \end{minipage}
	  \begin{minipage}[b]{0.5\linewidth}
		\centering
		\includegraphics[keepaspectratio, scale=0.55]{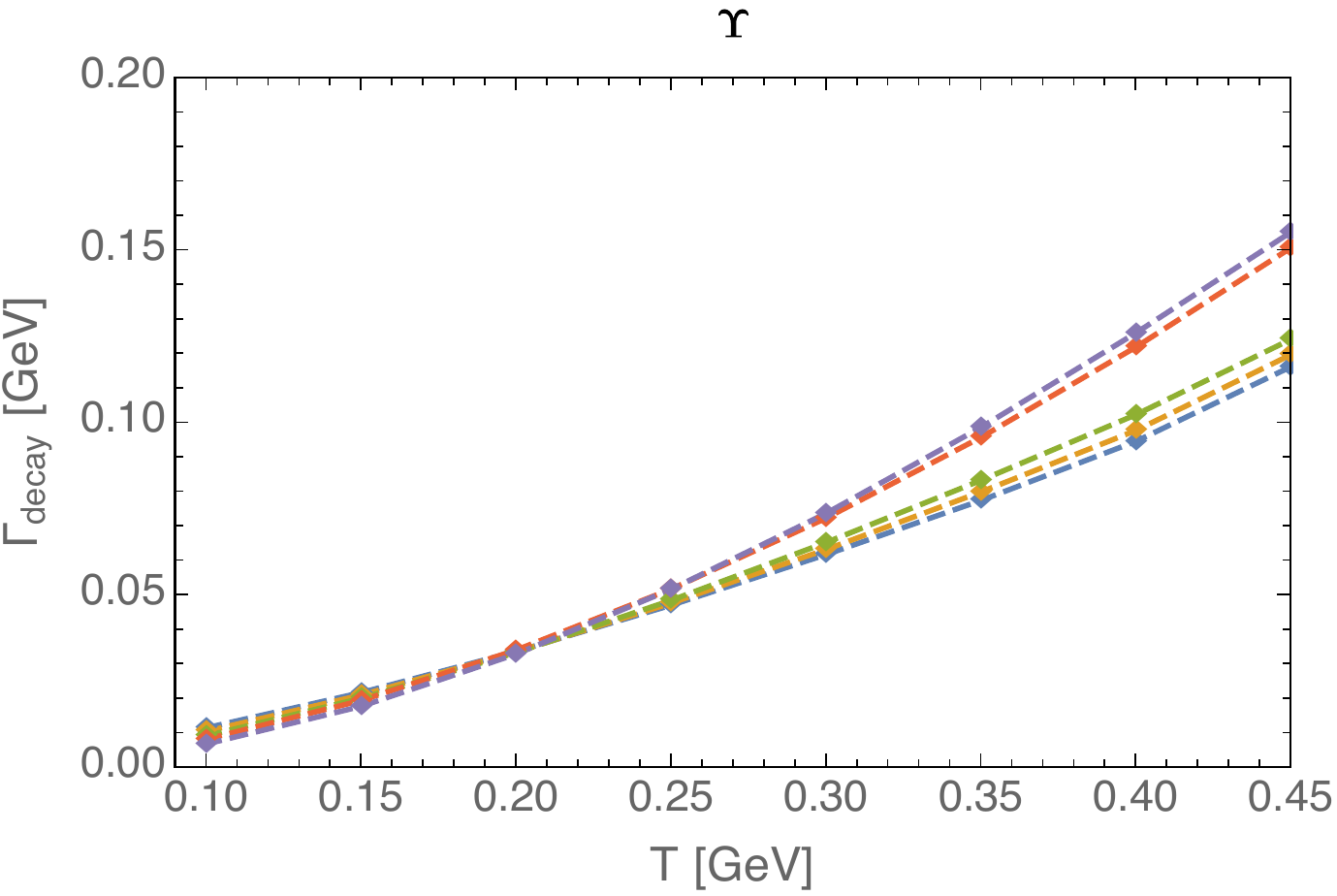}
	  \end{minipage} \\
	  \begin{minipage}[b]{0.5\linewidth}
		\centering
		\includegraphics[keepaspectratio, scale=0.55]{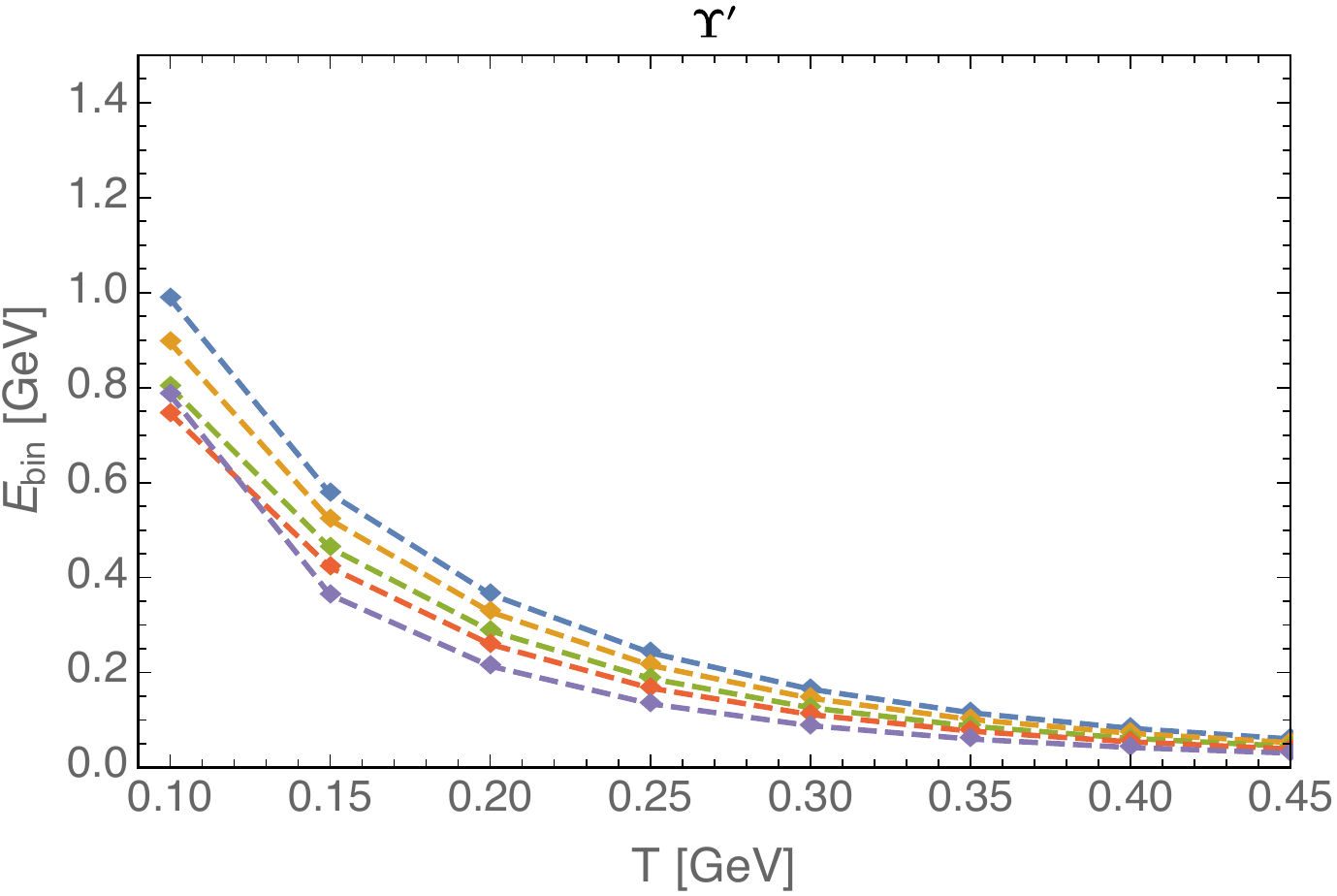}
	  \end{minipage}
	  \begin{minipage}[b]{0.5\linewidth}
		\centering
		\includegraphics[keepaspectratio, scale=0.55]{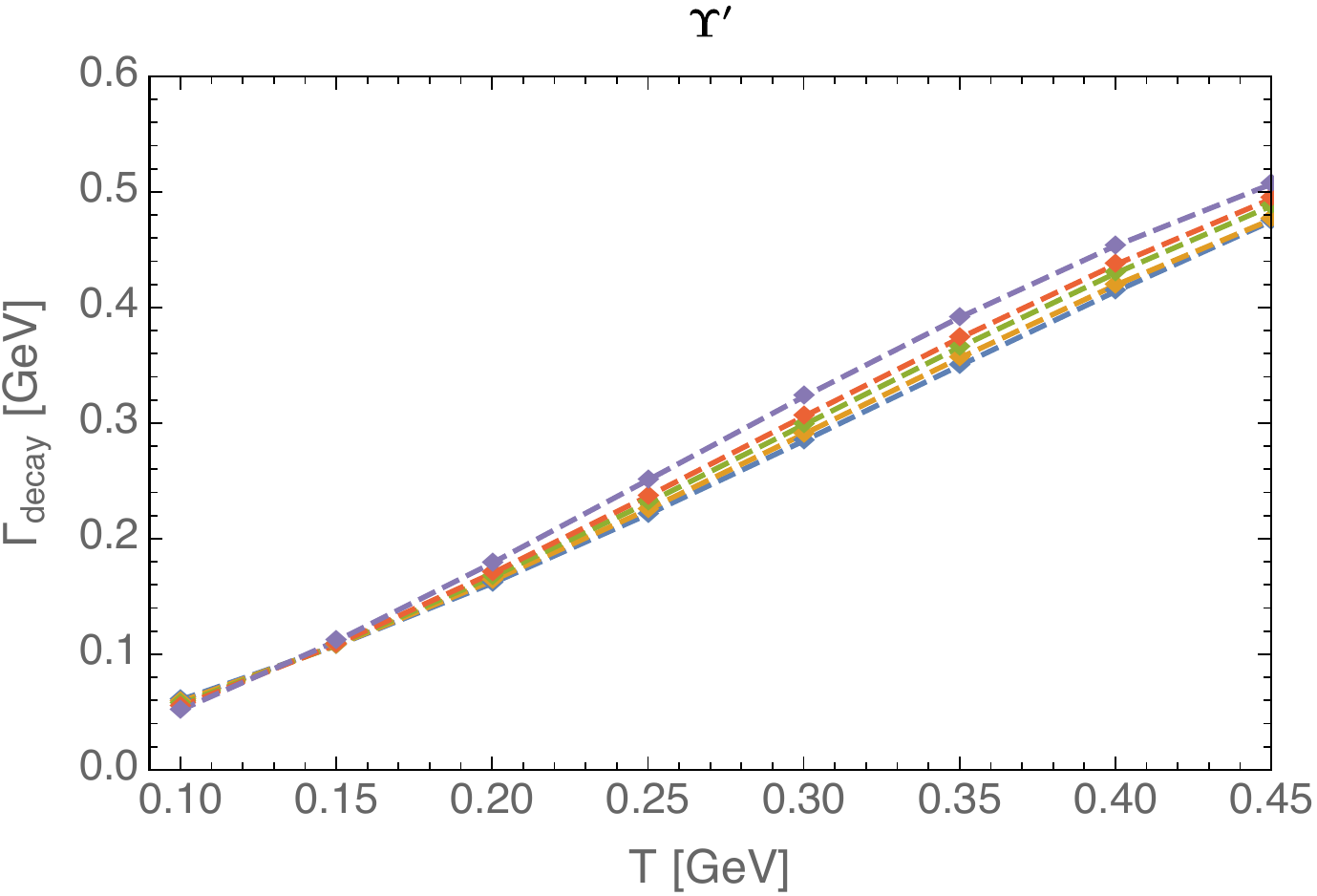}
	  \end{minipage} \\
	  \caption{
Binding energies (left column) and decay widths (right column)
for $J/\psi$ (top), $\Upsilon$ (middle), and $\Upsilon'$ (bottom) as a function of temperature. Different colors correspond to different values of the scale $\Lambda$. 
	  }\label{fig:e-b-lambda}
  \end{figure}

\begin{figure}[tb]
\centering
		\includegraphics[width=10cm]{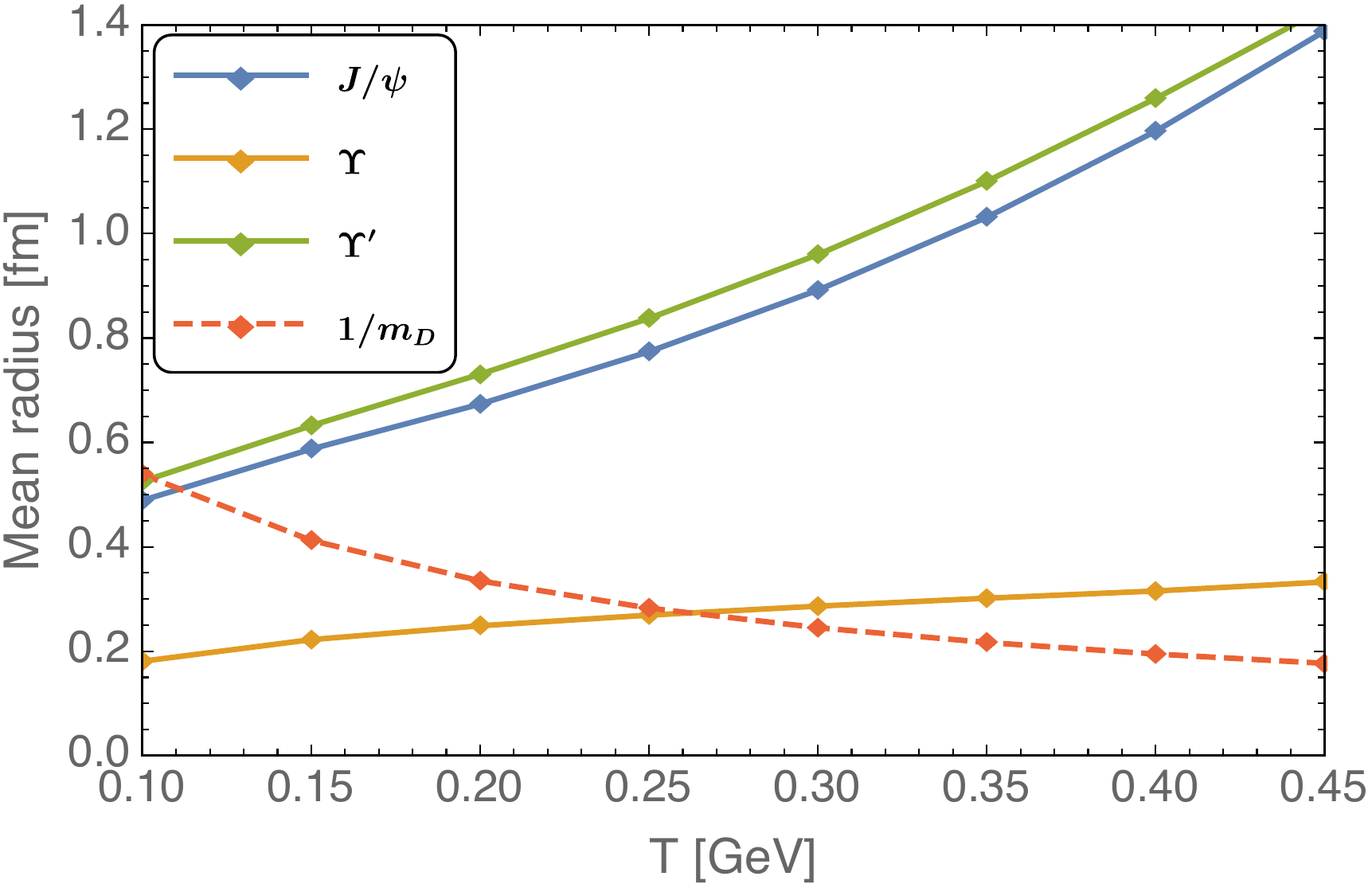}
\caption{
Mean radius of $J/\psi$, $\Upsilon$, and $\Upsilon^\prime$ compared with the Debye screening length $1/m_D$, as a function of temperature. 
}
	\label{fig:mean-r}
\end{figure}

In Fig.~\ref{fig:wf-T}, 
we show how the wave function is deformed 
as the temperature goes higher, for the ground state 
and the first excited states with $\ell=0$. 
A larger temperature results in a larger Debye mass, 
because of which the wave function is more delocalized
and the size of the quarkonium becomes larger. 
Since the bulk viscous correction comes
through the modification of the Debye mass, for 
the real part of the potential. 
When $\Phi>0$, the Debye mass becomes 
larger, and the wave function becomes more delocalized.

\subsection{Binding energies and decay widths }

\subsubsection{Dependence on the scale $\Lambda$}

Let us first consider the case without bulk viscous corrections.
Figure \ref{fig:e-b-lambda} shows the binding energies and decay widths for $J/\psi$, 
$\Upsilon$, and $\Upsilon'$ states.
As a check of the systematic dependence, we have changed 
the value of the scale $\Lambda$ by factors of $1/2$ to $2$, 
and different lines correspond to different values of $\Lambda$.
For $J/\psi$ and $\Upsilon'$, the binding energies decrease
and the decay widths increase for a larger $\Lambda$. 
$\Upsilon$ shows a different behavior:
the scale dependence changes around $T = 0.25\,{\rm GeV}$. 

The difference in behavior is related to the size of the wave function. 
In Fig.~\ref{fig:mean-r}, we show the mean radius of the quarkonia states, 
$\bar r \equiv \sqrt{\langle r^2 \rangle }$, with 
\begin{equation}
	\langle r^2 \rangle
	=
	\frac{
	 \int dr\, r^2 |\psi(r)|^2  r^2
	}{
	 \int dr \, r^2 |\psi(r)|^2 
	} . 
	\label{eq:meanr-sq}
\end{equation}
At around $T=0.25\,{\rm GeV}$, the mean radius of $\Upsilon$ becomes larger 
than the inverse Debye mass. 
When the wave function is small compared to the inverse Debye mass, 
the wave function is not sensitive to the screening and 
its behavior is determined by the Cornell-like potential. 
In the limit of tight wave function (large mass), 
the shape of the wave function is qualitatively close to the Coulomb wave function. Then, the binding energy is given by 
\begin{equation}
E_{\rm bin} \simeq \frac{ m_q \alpha^2 }{2} . 
\end{equation}
Since $\alpha$ is an increasing function of $\Lambda$, the binding energy increases 
for larger $\Lambda$, $\frac{\partial \alpha}{\partial \Lambda}>0$. 
On the other hand, at higher temperature, the size of the wave function grows and 
the Debye mass becomes relevant.
The potential asymptotically approaches a Coulomb potential with a different 
coefficient $\alpha' \equiv 2 \sigma / \widetilde{m}^2_{D, R}$. 
In this situation, the binding energy reads 
\begin{equation}
E_{\rm bin} \simeq  \frac{m_q (\alpha')^2}{2} = 
\frac{2 m_q \sigma^2}{m_D^4}
\propto \frac{1}{\alpha^2}. 
\end{equation}
Therefore, when the size of the wave function is comparable or larger than the Debye screening length, the binding energy is a decreasing function of $\Lambda$. 
This is why the $\Lambda$-dependence of
the binding energy changes around $T=0.25\,{\rm GeV}$ for $\Upsilon$. 
In the case of $J/\psi$ and $\Upsilon'$, the mean radius is larger than the inverse Debye mass, and the binding energy is always an increasing function of the scale $\Lambda$.

\begin{figure}[tb]
	\begin{minipage}[b]{0.5\linewidth}
	  \centering
	  \includegraphics[width=7.8cm]{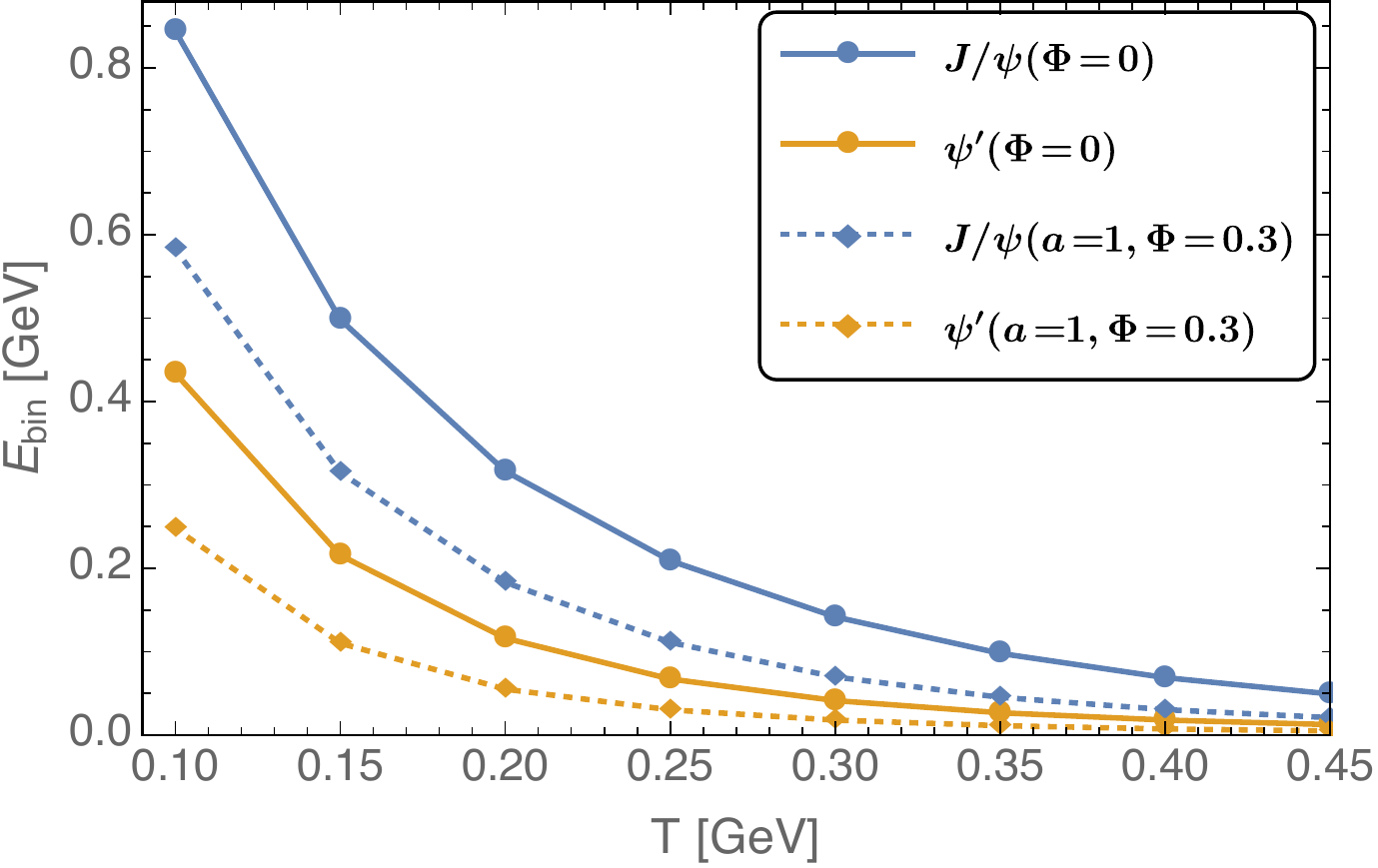}
	\end{minipage}
	\begin{minipage}[b]{0.5\linewidth}
	  \centering
	  \includegraphics[width=7.8cm]{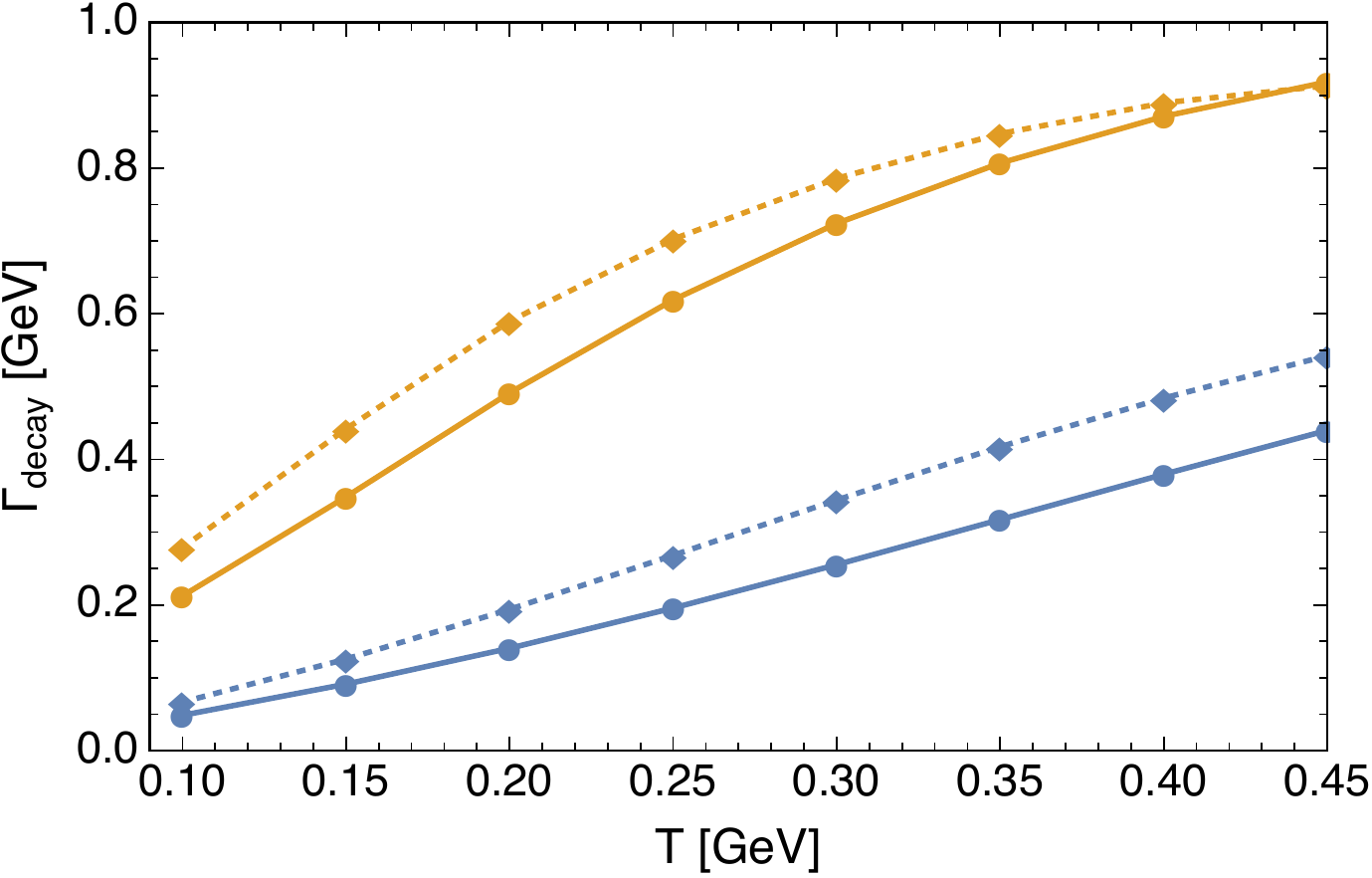}
	\end{minipage} \\
	\begin{minipage}[b]{0.5\linewidth}
		\centering
		\includegraphics[width=7.8cm]{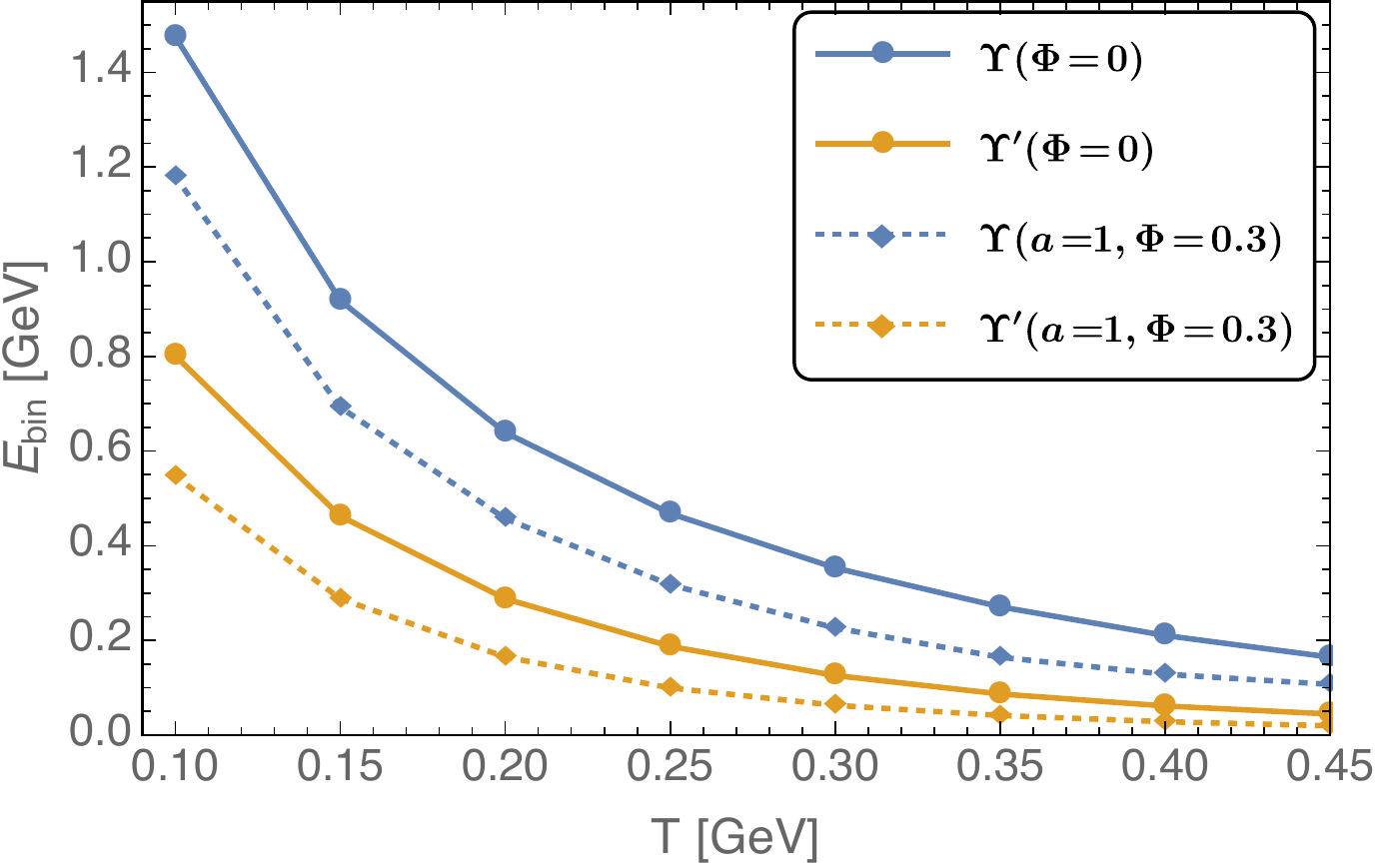}
	  \end{minipage}
	  \begin{minipage}[b]{0.5\linewidth}
		\centering
		\includegraphics[width=7.8cm]{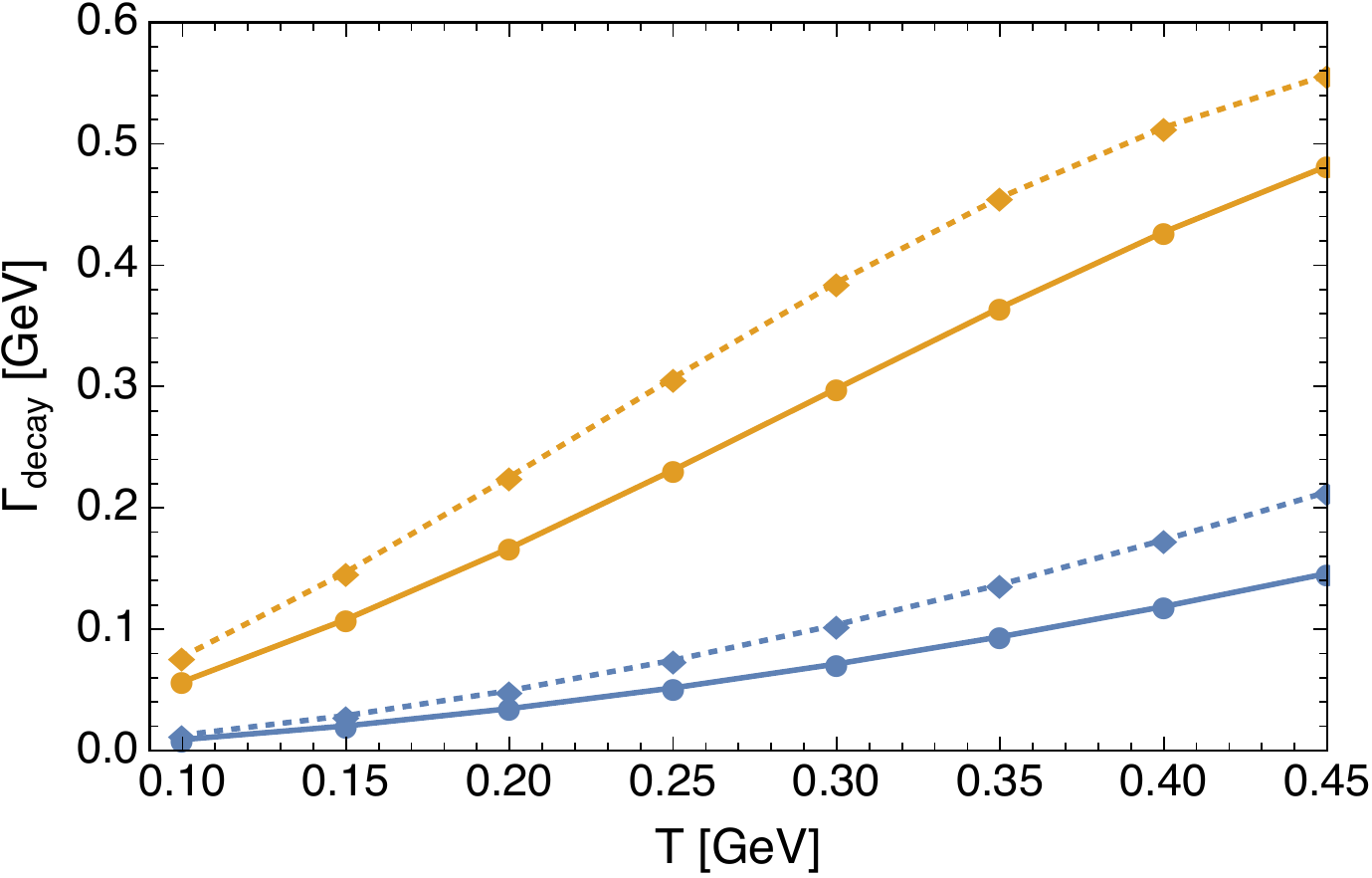}
	  \end{minipage} 
	  \caption{
		Modification of the binding energies (left)
		and decay widths (right) for 
		charmonium (top) and bottomonium (bottom) states 
		due to the bulk viscous corrections. 
		Solid lines are those without the bulk corrections 
		and dotted lines are with the modified ones. 
	  }\label{fig:e-gamma-bulk}
  \end{figure}

\begin{figure}[tb]
\centering
		\includegraphics[width=8.5cm]{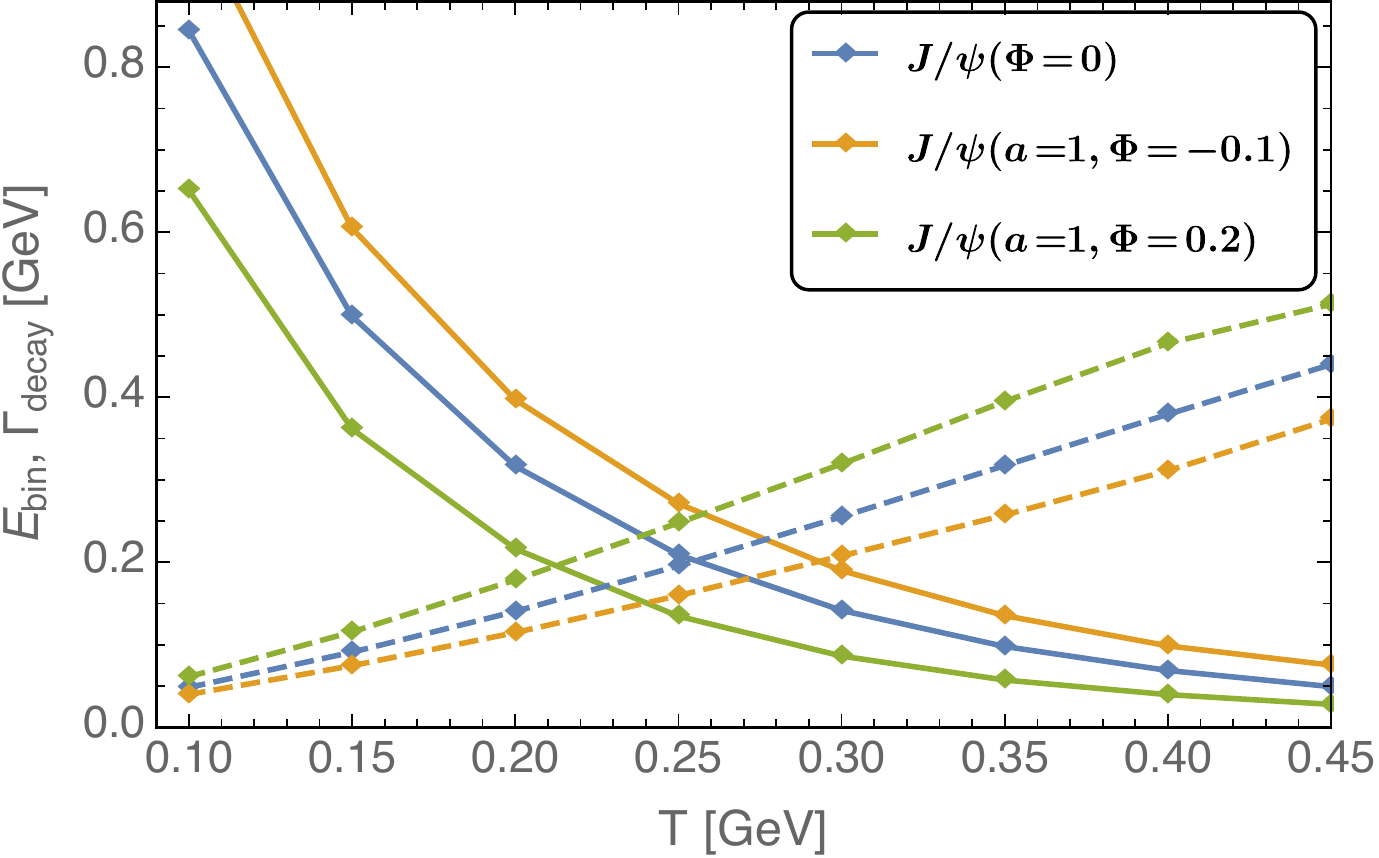}
\caption{
Shift of the melting temperature. 
The temperature dependence of the binding energies (solid lines)
and decay widths (dashed lines) of $J/\psi$ for different parameters $\Phi$ (indicated by different colors) is shown. 
The position of the intersection of a solid and dashed lines of the same color indicate the melting temperature for the corresponding parameter $\Phi$. 
}
\label{fig:jpsi-e-gamma}
\end{figure}

\subsubsection{Effect of bulk viscous corrections}

Now let us discuss the effect of the bulk viscous corrections 
on the binding energies and decay widths. 
In Fig.~\ref{fig:e-gamma-bulk}, 
we present the binding energies and decay widths 
for charmonium (top)  and bottomonium (bottom) states 
as a function of temperature. 
The solid lines correspond to the cases without bulk correction, 
and the dotted lines are with the bulk correction. 
Generically, the bulk correction for $\Phi>0$ lowers the binding energies, 
since the bulk viscous correction make the Debye mass heavier
and the screening becomes stronger.
For a negative $\Phi$, the effect is opposite. 

As for the decay widths, as 
shown in the right column of Fig.~\ref{fig:e-gamma-bulk}, 
the magnitude of the decay width is
enhanced in the presence of bulk viscous corrections. 
The decay width of the $\psi'$ state with the bulk correction 
(top-right) shows flattening at higher temperatures around $0.4\,{\rm GeV}$. 
This is related to the nature of the imaginary part
discussed in Sec.~\ref{sec:tkp}. 
As we discussed there, 
in the presence of the  bulk viscous correction, 
$|\Im V|$ is enhanced at small $r$, 
while it is suppressed at higher $r$. 
The wave function of $J/\psi$ is smaller in size, 
and in this region $|\Im V|$ is increased, 
resulting in the larger decay width. 
On the other hand, the excited states are larger and 
are more sensitive to the large $r$ part of $|\Im V|$. 
The enhancement of the decay width of $\psi'$ become 
saturated, because the wave function is now in 
the region where $|\Im V|$ is suppressed by  the bulk correction.

\begin{figure}[tb]
	\subfigure{
		\hspace{-8mm}
		\includegraphics[width=7.8cm]{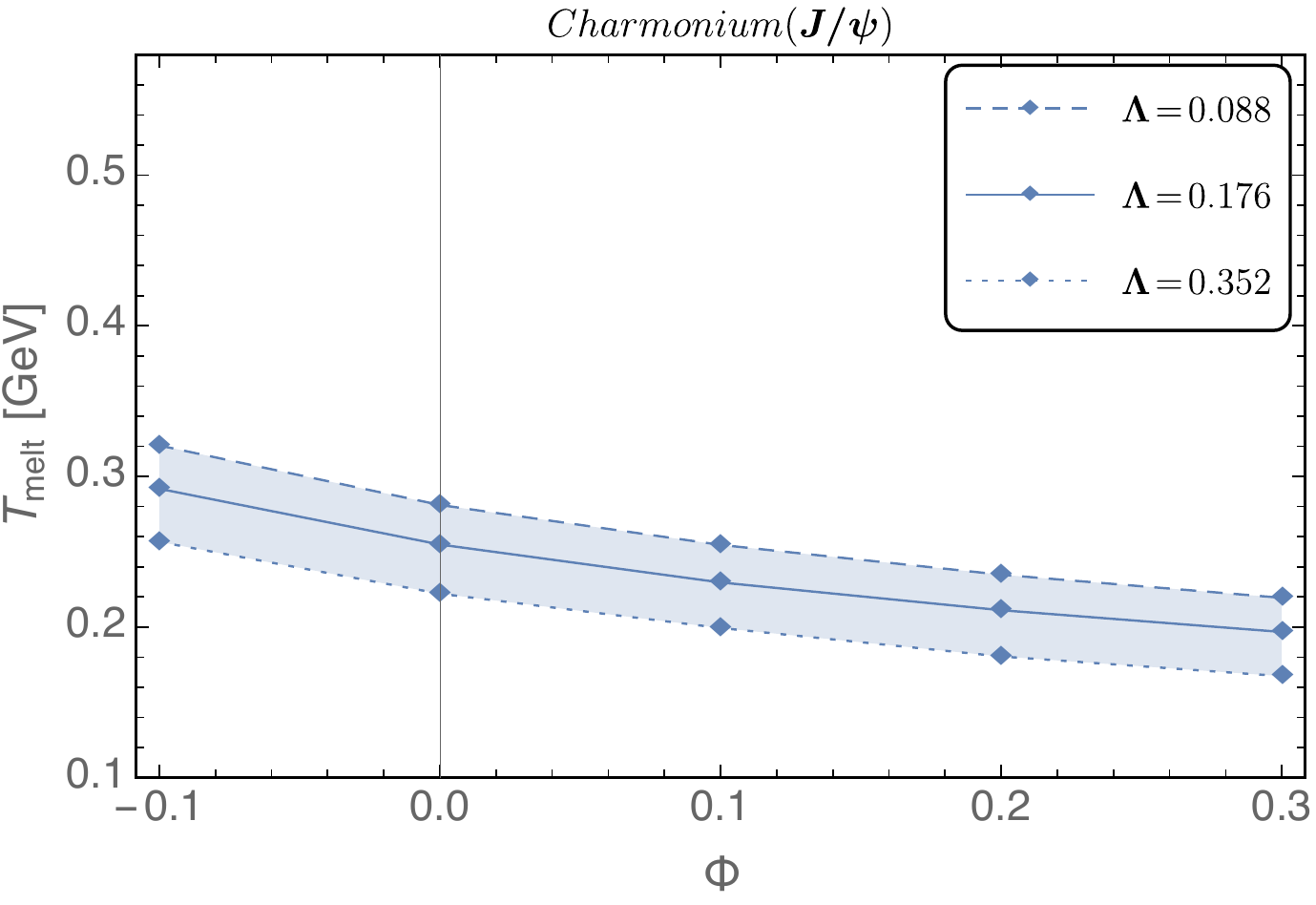}
		}
	\subfigure{
		\includegraphics[width=7.8cm]{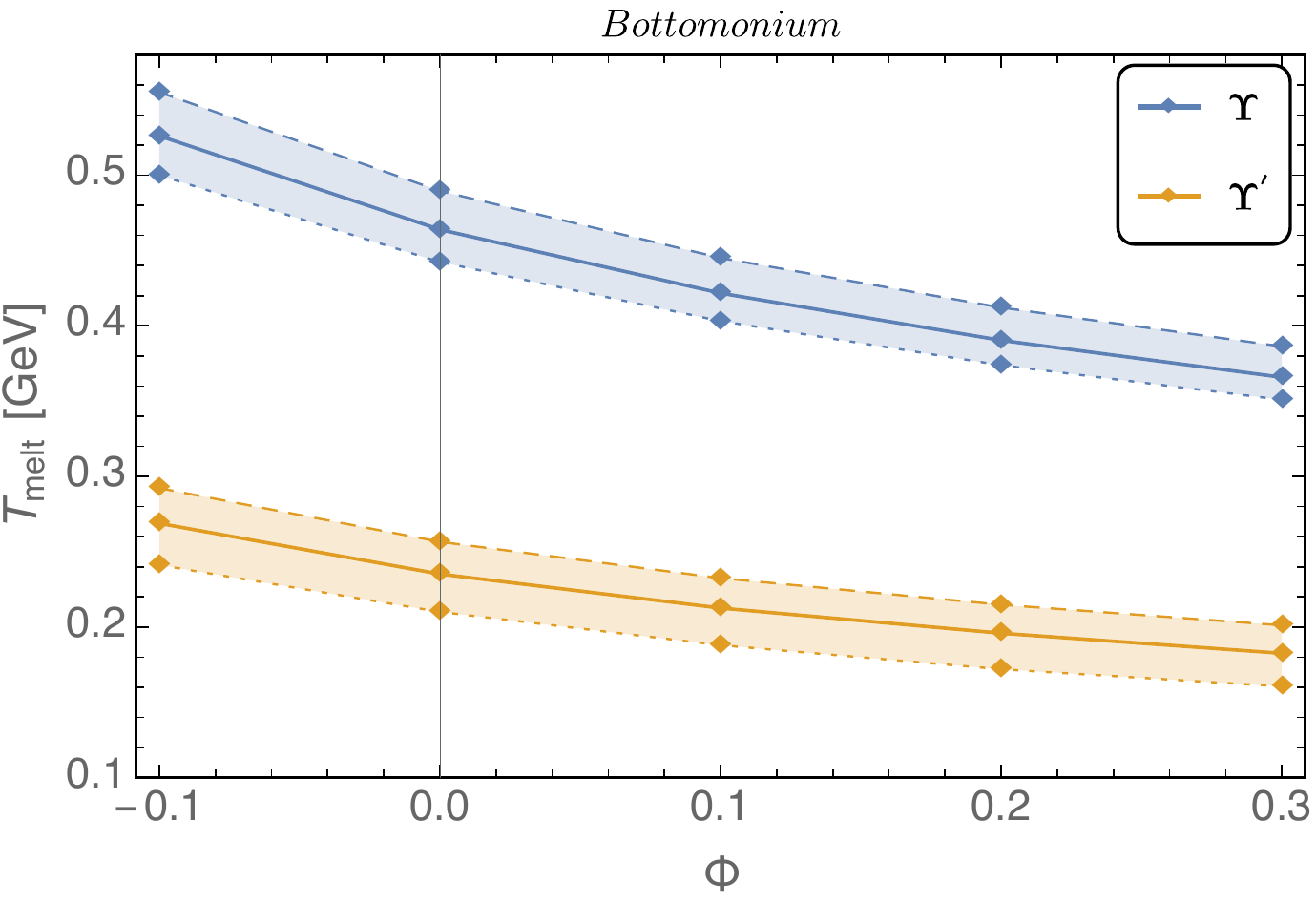}
		}
\caption{
Melting temperature
of $J/\psi$ (left), $\Upsilon$ and $\Upsilon'$ (right) 
as a function of the bulk viscous parameter $\Phi$ with $a=1$. 
The solid lines corresponds to $\Lambda = 0.176\, {\rm GeV}$,
around which the value of $\Lambda$ is varied
by a factor of $0.5$ ($\Lambda = 0.5 \times 0.176 \, {\rm GeV}$, dashed line), 
and a factor of $2$ ($\Lambda = 2 \times 0.176 \, {\rm GeV}$, dotted line). 
}
\label{fig:t-melt-c-b}
\end{figure}

Based on the computations of the binding energies and decay widths, 
we can make an estimate of the melting  temperature $T_{\rm melt}$ of 
quarkonium states. 
We adopt a common criterion
that the binding energy coincide with the decay width, 
 $E_{\rm bin}(T_{\rm melt}) = \Gamma(T_{\rm melt})$. 
For example, in Fig.~\ref{fig:jpsi-e-gamma}, 
we plot the 
$E_{\rm bin}$ (solid line) and $\Gamma$ (dashed line) of $J/\psi$ states. 
Different color corresponds to different parameter $\Phi$. 
The position at which a solid and dashed lines of the same color intersect is the melting temperature for the parameter $\Phi$. 
In Table~\ref{tab:meltingT}, 
we compare the 
melting temperatures computed in the current setup
(without bulk correction) 
as well as the results from Ref.~\cite{Lafferty:2019jpr}, 
that are based on the extraction of the potential with lattice QCD data
and subsequent computation of spectral functions. 
In order to gain a sense of uncertainty, 
we varied the scale $\Lambda$ around $0.176\,{\rm GeV}$, 
and the upper and lower numbers for our data in the table 
correspond to 
$\Lambda = 1/2 \times 0.176 \, {\rm GeV}$ 
and 
$\Lambda = 2 \times 0.176 \, {\rm GeV}$, respectively. 
\begin{table}
	\centering
	\begin{tabular}{ c  c   c }
		\hline
		\hline
		 & $\quad T_{\rm melt}$ [GeV]  $\quad$ &
	$\quad\quad$ Ref.~\cite{Lafferty:2019jpr}$\quad\quad$\\
		\hline
		$J/\psi$              &
		 $0.254^{+0.027}_{-0.032}$  &$0.267^{+0.033}_{-0.036}$\\
		$\Upsilon$        
		& $0.464^{+0.026}_{-0.022}$   &$0.440^{+0.080}_{-0.055}$\\
$\Upsilon^\prime$  
& $0.235^{+0.025}_{-0.022}$  & $0.250^{+0.050}_{-0.053}$\\
		\hline\hline
	\end{tabular} 	
\caption{
Melting temperatures for different states
in the absence of bulk viscous correction ($\Phi=0$) , 
compared with a recent work based on the lattice QCD~\cite{Lafferty:2019jpr}. 
}
\label{tab:meltingT}
\end{table}
Although 
there is essentially only one parameter $\sigma$ 
(the coupling $\alpha_S$ is given by the  one-loop result
and the Debye mass is given by the HTL) 
for the case of no bulk viscous correction, 
the computed melting temperatures is in a reasonable agreement. 

In Fig.~\ref{fig:t-melt-c-b}, we plot the melting temperatures 
of the states $J/\psi$, $\Upsilon$ and $\Upsilon'$
as a function of the bulk viscous parameter $\Phi$. 
The colored regions indicate 
the range between the values of $\Lambda$, 
$2 \times 0.176\,{\rm GeV}$ and $0.5 \times 0.176\,{\rm GeV}$, 
and the solid line is for $\Lambda = 0.176\,{\rm GeV}$. 
We observe a mild decrease of the melting temperature
as a function of $\Phi$. 
The slope of $\Upsilon^\prime$ is steeper 
compared to $J/\psi$ and $\Upsilon$. 
Therefore, we have found that the bulk viscous effect indeed 
affect the melting temperature, especially for excited states. 
When the plasma is close to the critical point, 
the bulk viscous contribution is expected to be enlarged. 
This can lead an anomalous behavior of observables 
related to heavy quarkonia in the beam energy scan program, 
although at this state it is difficult to make a quantitative estimate. 
In order to make a quantitative predictions, 
one should combine such behavior of quarkonia with 
a dynamical framework based on hydrodynamics.

\section{Summary and discussions}\label{summary}

Heavy quarkonia are useful in probing the nature of the medium around them, through the modification of their properties. 
In this article, we have studied how the non-equilibrium 
bulk viscous corrections are imprinted in 
the properties of heavy quarkonia. 
The bulk viscous correction modifies the distribution functions of thermal particles, from which the modified dielectric permittivity
is computed in the HTL approximation. 
Using the dielectric permittivity, we computed the 
heavy quarkonia potential and examined 
how the potential is  deformed in the presence of bulk viscous corrections. 
As for the real part of the potential, 
the bulk viscous correction parametrized by 
$\Phi>0$ leads to a larger Debye mass, hence stronger screening. 
The magnitude of the imaginary part is 
enhanced at short distances, and is suppressed at large $r$, 
as shown in Fig.~\ref{fig:imVTotal}. 
We have tried three prescriptions to accommodate the non-perturbative string-like 
potential at finite temperatures. 
Qualitative features of how the bulk viscosity affect the potential 
is found to be the same among different prescriptions, 
which indicates the robustness of the results.

We solved the Schr\"{o}dinger equation with the real part of the potential to obtain the deformed wave functions. 
We computed the binding energies and decay widths 
for $J/\psi$, $\psi'$, $\Upsilon$, $\Upsilon'$ states at 
different temperatures and the parameter $\Phi$ that quantify the bulk viscous corrections. 
Basically, a positive $\Phi$ leads to a Debye mass 
and it reduces the binding energy. 
On the other hand, the decay width is enhanced for a given temperature. 
Because of those, the melting temperature, 
at which the binding energy equals the decay width, is reduced 
for $\Phi>0$ and is enhanced for $\Phi<0$. 
When the system is near the critical point, the bulk viscous effect is expected to 
be enhanced, that would affect the melting temperature. 
It would be interesting to find an anomalous behavior of 
observables such as $R_{AA}$ in the beam energy scan program. 

Finally, let us make several comments about the possible future directions.  
A unique ability of potential models is that it can be extended 
to non-equilibrium situations. 
It would be interesting to combine the method of potentials 
with the technique of the QCD sum rule \cite{Morita:2009qk,Morita:2007pt, Gubler:2011ua, Araki:2017ebb,Gubler:2018ctz}
to gain insight into the properties of out-of-equilibrium QCD. 

Technically and conceptually, in order to understand the 
dynamical evolution of heavy quarkonia as a quantum state, 
the analysis based on the theory of  open quantum systems
would be desirable, which is actively studied recently \cite{Young:2010jq,Akamatsu:2011se,Akamatsu:2014qsa,Blaizot:2015hya,DeBoni:2017ocl,Kajimoto:2017rel,Brambilla:2017zei,Akamatsu:2018xim}. 
It would be interesting to apply/derive such a framework to non-equilibrium environments which can have anisotropic noises and see how the time evolution of quarkonia is affected.

Although we have demonstrated that 
heavy quarkonia have a potential to be sensitive to 
the bulk viscous nature of the surrounding media, 
how to experimentally probe this is a nontrivial question. 
Due to the non-equilibrium correction, 
the fluctuation-dissipation theorem is violated,  
which leads to two different Debye masses for 
retarded and symmetric propagators.  
If we can make an estimate of those two mass scales independently, 
it will be possible to explore the non-equilibrium bulk viscous corrections, 
that are expected to be significant near the critical point.

\section{Acknowledgement}
L. T. and Y. H. was funded  by the Korean Ministry of Education, Science and Technology, Gyeongsangbuk-do and Pohang City  
at the Asia Pacific Center for Theoretical Physics (APCTP). N. H. was funded by Department of Atomic Energy (DAE), India via National Institute of Science Education and Research.


\providecommand{\href}[2]{#2}\begingroup\raggedright\endgroup

\end{document}